\documentclass[fleqn,usenatbib]{mnras}

\usepackage{mathptmx}

\usepackage[T1]{fontenc}

\DeclareRobustCommand{\VAN}[3]{#2}
\let\VANthebibliography\thebibliography
\def\thebibliography{\DeclareRobustCommand{\VAN}[3]{##3}\VANthebibliography}

\usepackage{graphicx}	
\usepackage{amsmath}	
\usepackage{amssymb}	
\usepackage{color}
\usepackage{soul}
\usepackage{lineno}

\title[DESI Reverberation Mapping Feasibility]{Stacked Reverberation Mapping of High Redshift Quasars in DESI. \\ I. Feasibility Analysis}

\author[R. Alfarsy et al.]{Rahma Alfarsy,$^{1}$\thanks{E-mail: rahma.alfarsy@port.ac.uk}
R.E.A. Canning,$^{1}$
Eva-Maria Mueller,$^{2}$
Jessica Aguilar,$^{3}$
Steven Ahlen,$^{4}$
\newauthor
David Alexander,$^{5}$
Davide Bianchi,$^{6}$
David Brooks,$^{7}$
Peter Clark,$^{8}$
Todd Claybaugh,$^{3}$
\newauthor
Andrei Cuceu,$^{3}$
Tamara Davis,$^{9}$
Axel de la Macorra,$^{10}$
Saisrinivas Dhavala,$^{1}$
Victoria A. Fawcett,$^{11}$
\newauthor
Benjamin Floyd,$^{1}$
Andreu Font-Ribera,$^{12}$
Jaime Forero-Romero,$^{13}$
Enrique Gazta\~{n}aga,$^{1}$
Wei-Jian Guo,$^{14}$
\newauthor
Gaston Gutierrez,$^{15}$
Klaus Honscheid,$^{16}$
Richard Joyce,$^{17}$
Stephanie Juneau,$^{17}$
David Kirkby,$^{18}$
\newauthor
Theodore Kisner,$^{3}$
Anthony Kremin,$^{3}$
Claire Lamman,$^{19}$
Martin Landriau, $^{3}$
Laurent Le Guillou,$^{20}$
\newauthor
Paul Martini,$^{16}$
Hugh McDougall,$^{9}$
Aaron Meisner,$^{17}$
Ramon Miquel,$^{12}$
John Moustakas,$^{21}$
\newauthor
Seshadri Nadathur,$^{1}$
Nathalie Palanque-Delabrouille,$^{3}$
Zhiwei Pan,$^{22}$
Swayamtrupta Panda,$^{23}$
\newauthor
Ignasi~P\'erez-R\`afols,$^{24}$
Francisco Prada,$^{25}$
Ragadeepika Pucha,$^{26}$
Graziano Rossi,$^{27}$
Eusebio Sanchez,$^{28}$
\newauthor
David Schlegel,$^{3}$
Michael Schubnell,$^{29}$
Tom Shanks,$^{5}$
Ma{\l}gorzata Siudek,$^{30}$
David Sprayberry,$^{17}$
\newauthor
Gregory~Tarl\'{e},$^{29}$
Benjamin Alan Weaver,$^{17}$
and Hu Zou$^{31}$
\\
\\
Affiliations are listed at the end of the paper
}
\date{Accepted 2026 June 22. Received 2026 June 19; in original form 2025 November 21}

\pubyear{2025}

\begin{document}
\label{firstpage}
\pagerange{\pageref{firstpage}--\pageref{lastpage}}
\maketitle
\begin{abstract}
The broad line region of quasars has long been probed by reverberation mapping techniques that measure time lags between continuum and broad emission line variations. Stacked reverberation mapping has been proposed as a less observationally expensive alternative to traditional methods. This ensemble approach also reduces biases from small-number statistics. 
The Dark Energy Spectroscopic Instrument (DESI) is conducting the most extensive spectroscopic survey of quasars to date. We create mock light curves emulating expected DESI quasar observations at redshifts $1.48<z<5.2$ and luminosities $ 44.68 \leq \log \lambda  L_{1350 \AA{}} / \mathrm{erg\,s^{-1}} \leq 45.99 $ to test stacked reverberation mapping feasibility using sparse spectroscopic data paired with well-sampled photometric data. The pipeline, using the lag estimation code {\sc javelin}, successfully recovers the simulated C~{\sc iv} lags within one sigma of the true values using spectroscopic light curves composed of only a few spectral epochs (2-10) with irregular cadences. We investigate how observational factors, including C~{\sc iv} flux error magnitude, number of stacked quasars, and spectral epoch count, affect performance. This work motivates a pathway for future stacked reverberation mapping projects with large scale spectroscopic surveys of quasars having $\geq 2$  spectroscopic observations. Our results suggest an economical alternative for constraining and extending the radius-luminosity relation to higher redshifts and luminosities. Subsequently, this relation can be employed more reliably in single-epoch black hole mass measurements and quasar cosmology in these distant regimes.
\end{abstract}


\section{Introduction}

Extremely luminous active galactic nuclei (AGN) known as quasars are powered by the transfer of gravitational potential energy into radiation as material is accreted onto a central supermassive black hole (SMBH). The instabilities in the accretion process are observed as variations in luminosity across all wavelengths (e.g., \citealt{VandenBerk2004}). The source of this variability is still under discussion, but the accepted understanding is that it is due to a fluctuating accretion rate, amongst other factors \citep{Ulrich1999, Czerny1999, DeCicco2022, Wu2024}. The ionizing continuum emission sourced in the accretion disc is reprocessed by the orbiting dense clouds of gas in the broad line region (BLR) which appears as broad emission lines in the quasar's spectrum due to their high-velocities.

The compact structure of a quasar makes it difficult to resolve with imaging techniques. 
Thus, Reverberation Mapping (RM) has been developed to resolve cores in the time domain instead, enabling investigation of the dynamics and physical conditions of these regions. This is done by exploiting the variability of AGN, traced out by constructing light curves of the continuum and the broad emission lines. Since the latter are but reverberations responding to the former we can deduce a correlation and a time lag, $\tau$, between them. Measuring this lag allows us to directly infer the mean physical distance, symbolized by $R_{\rm{BLR}} \sim c \tau$, which the light travelled during the average time $\tau$ between the continuum emitting accretion disc and the photoionised clouds in the BLR \citep{Bahcall1972, Blandford1982}. There are several assumptions implied by this simple equation, including: that the continuum we observe is from a single point source and is directly related to the ionizing radiation, the gas properties and photoionisation models (e.g., \citealt{Panda2022}), that the light travel time is the dominant timescale contributing to the time lag \citep{Peterson1993}, and that the standard AGN model applies to every system we observe \citep{Rosborough2023}.

In practical terms, RM is the derivation, using mathematical or computational methods, of the transfer function that transforms the observed continuum light curve to the observed emission line light curve. This transfer function, often denoted as $\Psi$, contains information about the BLR's geometry, kinematics, and inclination angle \citep{Horne2003, Collin2006, Pancoast2011, Pancoast2014} and several forms and parameterisations have been investigated e.g., top-hat, Gaussian, and exponential (e.g., \citealt{Li2016}). The main parameter of the transfer function is the time lag, and the majority of studies have focussed on measuring this.

In the past, RM has been successful when monitoring AGN on a source-by-source basis which is observationally expensive \citep{Horne2002}. These studies target low redshifts and low luminosities using low ionisation lines such as H$\beta$ \citep{Denney2010, Rafter2013,Barth2015, Hu2015, Du2016}. RM of high redshift and high luminosity quasars with high ionisation lines such as C~{\sc iv}$\lambda$1549\AA{} has been attempted too \citep{ Trevese2014,Jiang2016, Kaspi2021} but retrieving a robust lag measurement has been more difficult and rarer. C~{\sc iv} is a UV line which, when observed with ground-based telescopes, can only be seen in high redshift quasars. Such quasars require extensive observational campaigns in order to measure a time lag since the lag is cosmologically time dilated. Although it has been found that high ionisation lines have shorter rest-frame lags than low ionisation lines due to stratification of the BLR \citep{Horne2020}, when observing them in high redshift sources their redshifted observed frame lag is longer than the observed frame lag of low ionisation lines found in low redshift sources \citep{Kaspi2007}. This means that RM of high ionisation lines is limited by the requirement of longer observation programs to measure a reliable lag.

To date, robust RM lag measurements have been made for dozens of quasars leading to the finding of a tight correlation between $R_{\rm{BLR}}$ and the monochromatic luminosity of the quasar, the $R$-$L$ relation, in the form:
\begin{equation}
\label{eqn:rl}
    \log(R_{\rm BLR}) = a \log(L_{\rm cont}) + b
\end{equation}
where a and b are constants (e.g., \citealt{Kaspi2005,Bentz2013}).
The $R$-$L$ relation is used to calibrate mass-scaling relations to estimate single-epoch BH masses (e.g., \citealt{Kollmeier2006, Pan2025}). It also has potential for transforming quasars into standardisable candles as an upcoming probe for cosmology \citep{Watson2011, Panda2019, MartinezAldama2019, Czerny2023b, Jaiswal2024}.
For any given line, we need a statistical sample of lag measurements to constrain the $R$-$L$ relation. The usage of this relation requires further investigation with a larger and more diverse sample of quasars.

In an effort to achieve this, RM schemes have been constructed by both the Sloan Digital Sky Survey Reverberation Mapping program (SDSS-RM) and the Australian Dark Energy Survey Reverberation Mapping project (OzDES-RM) to produce a statistical sample of time lags. The observation of hundreds of AGN within these multi-object RM surveys has significantly contributed to expanding the previously small RM populations and advancing our understanding of quasar properties \citep{King2015, Shen2015, Grier2019,Hoormann2019,Homayouni2020,Yu2021,Malik2023}. This research has paved the way for the development of an array of time lag measurement techniques. These include techniques that 1) do not assume a parametric form for the transfer function $\Psi$ nor a model for variability, such as cross-correlation function analyses, and 2) those that do assume a parametric form for $\Psi$ as well as a damped random walk (DRW) model for continuum variability such as Bayesian Markov Chain Monte Carlo (MCMC) codes. However, limitations still lie in low signal-to-noise data, intensive observations still yielding limited sampling of light curves \citep{Malik2022}, and challenges in recovering reliable time lag measurements \citep{Li2019, Penton2022}, all of which are more profound for high ionisation lines like C~{\sc iv} which require long-term observational campaigns. 

Photometric RM has been conducted successfully to measure accretion disc lags \citep{Haas2011, Yu2020b, Kovacevic2022, PozoNunez2023}. Photometric RM of the BLR has also been attempted as a less expensive option for obtaining emission light curves than spectroscopy  \citep{Zu2016} and is a promising tool considering surveys such as LSST are underway \citep{Ivezic2019}. For example, \citealt{Panda2024} demonstrated the possibility of bypassing the difficulties of high redshift RM by using scaled photometric accretion disc lags. Yet, methods to retrieve accurate lags from photometry are still less developed \citep{Read2020}.

An attempt to circumvent insufficient spectroscopic light curves has been sought in Stacked Reverberation Mapping proposed by \citet{Fine2012} and confirmed by \citet{Fine2013}. By combining the time lags measured for individual quasars with similar physical properties, an average time lag with a higher statistical significance can be calculated. This boost in signal-to-noise ratio (SNR) of the time lag is achieved by leveraging data in a way that traditional individual measurements cannot, as demonstrated in both SDSS-RM \citep{Li2018} and OzDES-RM \citep{Malik2024} samples. Stacked RM does not aim to improve upon single source time lags but it is a statistical technique that aims to provide reliable lags for quasar populations for which traditional RM is observationally expensive and challenging for faint sources, restricting the probed luminosity-redshift-BH mass plane. The most constrained $R$-$L$ relations have been restricted to low redshift systems and they have been used to extrapolate the $R$-$L$ relation to high redshift systems for single-epoch BH mass estimation; the assumptions comprised in doing this may not be valid and may lead to incorrect BH mass measurements.

 
The advent of Stage IV spectroscopic surveys presents an exciting frontier in quasar RM research. DESI and its prospective successor, DESI-II, hold promise in broadening the scope and scale of quasar RM investigations. DESI's unprecedented statistical sample of quasars may enable the development of large scale stacked RM that is sensitive to quasar diversity. 

In this paper, we outline a method for stacked RM with large spectroscopic surveys and test its feasibility on simulations by demonstrating a developed pipeline based on the first year of data from DESI. We combine the emulated spectroscopic emission line measurements expected from DESI with continuum photometry imitating Zwicky Transient Facility (ZTF) cadence. The pipeline presented explores a novel procedure for stacked RM.

For the purposes of this project, we focus on RM with the broad C~{\sc iv} emission line. It is one of the stronger lines in this regime observed to high redshifts. We are motivated by the poor characterisation of the $R$-$L$ relation by high redshift RM thus far, as well as its use for high redshift quasar cosmology. With traditional RM, multiyear baselines are needed for C~{\sc iv} light curves which will inevitably be plagued by seasonal gaps weakening the potency of single-object RM. What's more, C~{\sc iv} flux measurements are notoriously challenging due to the prevalence of an outflow component creating an asymmetrical non-BLR component and the presence of an Fe {\sc ii} pseudo-continuum underlying the broad emission line that require additional efforts to account for. Subsequently, there are currently only of the order of tens of C~{\sc iv} lags measured by previous studies \citep{Peterson2005, Metzroth2006, Kaspi2007, Trevese2014, Derosa2015, Lira2018, Shen2019, Hoormann2019, Grier2019, Kaspi2021, Penton2022, Shen2024}. 

We aim to assess whether the high luminosity-high redshift regime with slow variability patterns and expected C~{\sc iv} lags on the order of several months to a year can be reliably accessed with stacked RM by constructing mock quasar light curves with known lags. In turn, we investigate what the limitations of the method are and suggest an observational strategy accordingly. This inaugural paper sets up the RM technique that will be applied to DESI data in a forthcoming paper. Two further studies will also complement this project. The first is a characterisation of the C~{\sc iv} line for DESI quasars and an examination of what subset of the quasar population is suitable for C~{\sc iv} RM. The second is employing this RM pipeline without the assumption of a cosmology. By deducing cosmology simultaneously with the $R$-$L$ relation, we aim to wield quasars as a direct probe of cosmology, focusing on the nature of Dark Energy.

This paper is organized as follows. In the next section, we provide a basic introduction of the physics behind stacked RM. Next, we describe the DESI survey and what it has to offer towards RM studies in Section \ref{sec:DESI}. In Section \ref{sec:phot}, we explore photometric surveys which could supplement the DESI data and we justify the decision to select the ZTF survey for simulating the photometric light curves. The generation of the mock light curve suite made to resemble the expected quasar population and data available for these two surveys is described in Section \ref{sec:mocks_method}. Section \ref{sec:lag_measurement} details the stacked RM pipeline for recovering time lags from the simulated light curve pairs making use of the time lag measurement code {\sc javelin}. The evaluation of the results of the pipeline as well as a rigorous investigation to determine its limitations are presented in Section \ref{sec:results}. A thorough discussion of the use of this method can be found in Section \ref{sec:discussion}. The feasibility of stacked RM is reinforced in Section \ref{sec:CCF} by providing an alternative pipeline using cross-correlation functions. Finally, we conclude our findings in Section \ref{sec:conc}.

Throughout the work in this paper we adopt a flat $\Lambda$CDM cosmology with $\Omega_\Lambda = 0.7, \Omega_M =0.3, H_0 = 70$ kms$^{-1}$Mpc$^{-1}$ when calculating luminosities. A cosmology-independent setup will be experimented in future work.


\section{Introduction to Stacked Reverberation Mapping}

The idea of utilising RM to indirectly probe the central engine of quasars was first proposed by \citet{Blandford1982} at a time when the variability of quasars in both continuum and emission lines was being discovered (also see earlier discussion in \citealt{Bahcall1972}). RM was first conducted on NGC 5548 \citep{Clavel1991, Peterson1991, Peterson1992, Dietrich1993, Reichert1994} which confirmed its potential and revealed certain intricacies of AGN structure. Essentially, RM is a means of side-stepping the hindrances of determining quasar structure in the spatial domain by instead exploiting the pervading spectral variability in the time domain to deduce spatial information indirectly. The micro-arcsecond angular resolution needed for a telescope to directly resolve the structure within the central inner parsec of quasars had previously not been achievable. This has now been accomplished by the GRAVITY instrument with which the results of RM have been independently confirmed \citep{Sturm2018, Lutz2024, Gravity2024a, Gravity2024b}. 

Broad emission lines are a defining characteristic of (Type I) quasar spectra, the Doppler broadening they display is a reflection of the orbital velocities of their emitting region, known as the BLR, around the central SMBH. The underpinning concept of RM is that the broad emission lines emitted by optically thick clouds in the BLR are the direct consequence of the reprocessing of the ionising continuum from the accretion disc. Hence, any variations seen in the flux of the broad emission lines are but an echo or reverberation of the flux variations of the ionising source. The graph that traces the variations of a particular wavelength of the continuum radiation or the flux of a particular species of broad emission line over time is known as a light curve. A comparison between the light curve of the continuum emitted from the accretion disc and the light curve of the broad emission line emitted from the BLR reveals a time delay between these two components and is known as the lag. The variations in the broad emission line flux lags behind the continuum because the ionizing continuum radiation had to travel the distance between the accretion disc and the BLR before it was reprocessed by the clouds therein. The light-travel time is reflected by a lag we can measure allowing us to deduce the radius at which the clouds are orbiting around the central SMBH by the simple relation:
\begin{equation}
    R_{\text{BLR}} = \tau c
\end{equation}
where $R_{\text{BLR}}$ is the radius of the broad line region, $\tau$ is the time lag found from comparing the two light curves, and $c$ is the speed of light.

The transfer function $\Psi(\tau)$ is the function that transfers the continuum light curve, $C(t)$, to the broad emission line light curve $L(t)$:
\begin{equation}
    \label{eqn:transfer}
    L(t) = \int^{\infty}_{-\infty} \Psi(\tau) C(t-\tau) d\tau.
\end{equation}
In essence, RM aims to extract the time lag of the transfer function which characterises the broad emission line response from our light curve observations (for RM overviews see \citealt{Netzer1997}; \citealt{Peterson2004}). 

A major problem in RM is how exactly should the comparison between the continuum and broad emission line light curves be done in order to measure a time lag. Several methods have been developed to measure the time lag which are described in Section \ref{sec:lag_measurement}. A long list of assumptions have gone into the theoretical set-up of RM, for example see \citet{Perterson1994}. Another set of assumptions are compounded depending on the method you use to measure the time lag.

Before the conception of stacked RM by \citet{Fine2012}, all methods of measuring the time lag had a high dependence upon well-sampled light curves for their success (see \citealt{Horne2002} for further detail). This requirement has limited the number of quasars that have undergone RM analysis due to the observational costs that each one incurs. Moreover, the variability of a quasar is inversely correlated to the quasar's luminosity \citep{Cristiani1997}. Deducing the time lag between two light curves of a highly variable source is more straightforward than that of a low variability source. This has meant that there has been a preference for conducting RM on a high variability and therefore low luminosity and low redshift quasar sample. In optical-UV spectra the broad emission lines that appear at low redshifts include H$\alpha$ and H$\beta$; most forerunning RM studies had a focus on these lines. Broad emission lines in higher redshift spectra such as Mg {\sc ii} $\lambda$2798\AA{ and C~{\sc iv} have recently also had an increase in successful RM studies (e.g., \citealt{Shen2024}).

Stacked RM has been developed with the aim to 1) expand the number of quasars included in RM studies by greatly reducing the observational cost, and 2) make RM accessible to the whole quasar population including high luminosity and high redshift quasars. The principle is to perform RM on a large ensemble of quasars instead of on an individual basis. Stacked RM gives us a window into the structure of quasar sub-populations as opposed to the structure of individual quasars. The requirement is that a quasar must have at least two spectroscopic observations to be included, i.e. only two data points are needed in the broad emission line light curve. A necessary assumption of the stacking process is that all of the quasars being stacked have similar BLR radii and hence similar lags. As seen in Equation \ref{eqn:rl}, the radius is dependent on the luminosity of the quasar and this relationship between them is tight. For example, the scatter around the H$\beta$ $R$-$L$ relation was previously found to be as low as 0.13~dex \citep{Bentz2013} and more recent studies have found a scatter of $\sim0.4$ dex \citep{Shen2024}. The quasars contributing to the C~{\sc iv} $R$-$L$ relation display a larger scatter \citep[$\sim 0.5$ dex,][]{Shen2024} and this spurs further motivation for exploring the intrinsic and non-intrinsic sources of scatter for C~{\sc iv}. Given little variation is found between the BLR radii of individual sources of a similar luminosity, it is therefore necessary to confine the sample of quasars being stacked to those of similar luminosity. The success of RM now depends on the curation of a large enough sample of quasars with similar BLR radii rather than intensive observations. 

Each pair of quasar light curves is processed through a time lag measurement method. Given that lag measurement methods require well-sampled light curves, the measurement from these sparse light curves can only be poorly constrained. Stacked RM differentiates itself from RM by an additional stacking process which is the combination of the quasar's lag measurements. Each individual lag measurement, although poor, contains a faint signal of the true lag. There is an assumption made that systematic errors in the signal of individual sources is completely suppressed in the stack. The stacking procedure reinforces the true lag signal with each additional quasar stacked and, with enough quasars, a reliable time lag can be obtained.

\section{DESI}\label{sec:DESI}
DESI \citep{Levi2013,DESI2016a,DESI2016b} is a multi-object spectrograph installed on the 4m Mayall Telescope at Kitt Peak National Observatory \citep{DESI2022}. It consists of 5000 fibers that are robotically positioned on targets covering a $\sim$8 deg$^2$ field \citep{Silber2023, Miller2024, Poppett2024}. The survey footprint will cover 14,000 deg$^2$ visible from the northern hemisphere. DESI was initially designed to be a 5-year survey but has now been extended to be continue for eight years. Survey Validation (SV) programs were conducted between December 2020 and May 2021 before the subsequent beginning of the Main survey. SV programs were conducted to refine target selection and so probed a 140 deg$^2$ field repeatedly over many nights to obtain deeper, high SNR, and more complete data than the Main survey \citep{Alexander2023}. The sample of quasars contained within it tend to have an above average number of epochs. SV data was released as part of the Early Data Release (EDR, \citealt{DESI2023a, DESI2023b}) and will continue to be included in future data releases. Year 1 (May 2021 - June 2022) of the main survey data has been released as Data Release 1 (DR1, \citealt{DESI2024I}).

The observed targets are categorized into the following classes: Milky Way Survey \citep{Allende2020}, Bright Galaxy Survey \citep{Ruiz-Macias2020}, Luminous Red Galaxies \citep{Zhou2020}, Emission Line Galaxies \citep{Raichoor2020}, and quasars \citep{Yeche2020}. There are additional secondary target classes for when there are spare fibers unable to point at a primary target due to the geometry of the robotic fiber positioners. DESI aims to observe three million quasars (around half of these are already observed in Year 1) quadrupling the known number of quasars prior to the survey. These quasars are targeted using the Legacy Imaging Survey DR9 \citep{Dey2019, Chaussidon2023} which was sensitive down to $r\sim23$. The DESI spectral range is 3600\AA{} - 9800\AA{} and has a medium spectral resolution of R $>2100$, R $>3200$ and R$>4100$ for the blue (3600 - 5900\AA), green (5660-7220\AA{}), and red (7470-9800\AA{}) cameras respectively \citep{DESI2022}. The data is processed using the spectroscopic reduction pipeline detailed in \citet{Guy2023}, which performs sky subtraction as well as flux and wavelength calibration, yielding reliable spectra for inter-epoch comparison. This is immediately followed by a template-fitting pipeline named Redrock providing a spectral classification and redshift estimate for each target on the night of its observation (Bailey et al. in prep, for the quasar templates see \citealt{Brodzeller2023}).

Amongst DESI's key aims is to use spectroscopic redshifts of each of the target classes to calculate high precision measurements of Baryon Acoustic Oscillations \citep{DESI2024a} and Redshift Space Distortions \citep{DESI202b} in order to parametrise the nature of Dark Energy. Yet, the quality and quantity of data DESI collects to achieve this will also provide a treasure trove for astrophysical studies. DESI observes quasars over a wide redshift range covering all broad emission lines typically used for RM such as H$\alpha$, H$\beta$, Mg {\sc ii}, and C~{\sc iv}. Figure \ref{fig:zregime} presents DESI's ability to access these broad lines over a wide redshift range including significant overlap between broad emission lines which enables direct comparison between RM of different emission lines. For example, DESI can supplement the study presented in this project on C~{\sc iv} lags by corroborating the lags and the method used to obtain them by repeating the pipeline on a statistically significant subset of our sample which also contains a Mg {\sc ii} broad emission line in order to evaluate the extent to which C~{\sc iv} lags are in agreement with Mg~{\sc ii} lags.

There are various Value Added Catalogues (VACs) that aid in exploring DESI data. FastSpecFit (Moustakas et al. in prep) is an extensive VAC that contains results of spectroscopic fitting of stellar continuum and emission-lines for all extra-galactic targets from the Redrock files. In the development of the pipeline, we use FastSpecFit VACs for both Main survey and Survey Validation programs (Iron VAC v2.1) as our parent catalogue for evaluating DESI's potential for RM.

\begin{figure}
\includegraphics[width=8cm]{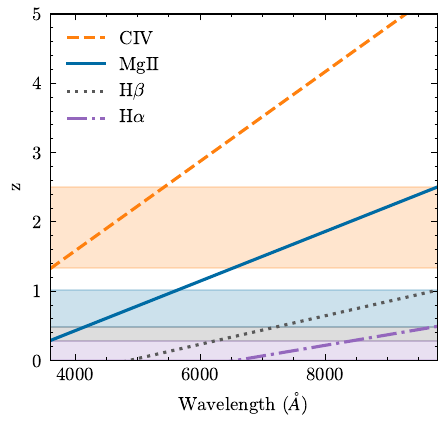} 
\caption{ Plot showing the wavelengths at which the typically brighter broad  emission lines found in the DESI spectrum H$\alpha$, H$\beta$, Mg {\sc ii}, and C~{\sc iv}  are situated in the spectrum at a given redshift. The DESI observed wavelength range is from 3600\AA\ to 9800\AA. These lines also indicate the overall redshift ($z$) ranges the broad lines are visible in the spectrum. The shaded regions highlight the redshift ranges in which more than one broad line is present in the spectrum (H$\alpha$ and H$\beta$ - purple, H$\alpha$ and H$\beta$ and Mg {\sc ii} - grey, H$\beta$ and Mg {\sc ii} -blue, and Mg {\sc ii} and C~{\sc iv} - orange).}
\label{fig:zregime}
\end{figure}

\section{Photometric data} \label{sec:phot}

The broad emission line light curve requires a spectrum to measure the emission line flux. On the other hand, the continuum light curve can be created from photometric observations. The continuum fluxes need to be dominated by the accretion disc radiation as opposed to the host galaxy's starlight. The selection of our C~{\sc iv} quasar sample means the photometric bands are probing the UV spectrum which is indeed dominated by the quasars luminosity and has minimal host galaxy contamination \citep{VandenBerk2004}.
Photometric surveys tend to have far more epochs of observations for their targets than spectroscopic surveys, which allows us to well characterize the continuum variations. So, we aim to pair the spectroscopic emission line data with well-sampled photometric light curves. 

We considered various surveys including DECaLS (DESI's imaging survey), Sloan Digital Sky survey, Panstarrs, DES, KiDS, Gaia, NeoWISE, and ASAS-SN. Our criteria for evaluation is sky footprint overlap with DESI, cadence of observations, and limiting magnitudes to access fainter objects at the high redshifts we are concerned about. In practice, we could utilise different surveys to cover all magnitude ranges and areas of the sky. 

For the study presented in this paper, we have opted to simulate data from the Zwicky Transient Facility (ZTF, \citealt{Bellm2019}), an optical time-domain survey using the Schmidt telescope at Palomar Observatory designed to measure temporal variability. It scans the northern sky (declination $>$-28) every $\sim$ 2 days in \textit{r}-band. ZTF's excellent cadence and overlap with the DESI footprint, captured in Figure \ref{fig:sky}, makes it an ideal survey to source photometric light curve counterparts to our sample for the continuum variability analysis. After accounting for ZTF's limiting magnitude, we expect at least 80\% of our sample to have suitable counterparts, the remaining mostly lost due to falling outside of the footprint or having too few points in the light curve (<100). The light curve data analysed in this work is taken from Data Release 23, which spans from March 2018 to October 2024. We assert the need for >100 epochs per object as a proxy for having sufficient observations stretched quasi-uniformly over a long baseline (at least a few times the lag) for reliable lag estimation (e.g., \citealt{King2015}). ZTF started operating in March 2018 meaning there will be light curve data points preceding the first DESI observations far beyond the longest lags we expect to detect, ensuring that the continuum and emission line flux variations will overlap. Furthermore, the ZTF survey will continue observing throughout the rest of the DESI survey timeline. ZTF \textit{gri} bands have detection limits of \textit{g} $\sim$ 20.8, and \textit{r} $\sim$ 20.6 mag (AB, 5$\sigma$) \citep{Masci2019}.

\begin{figure*}
\includegraphics[width=\textwidth]{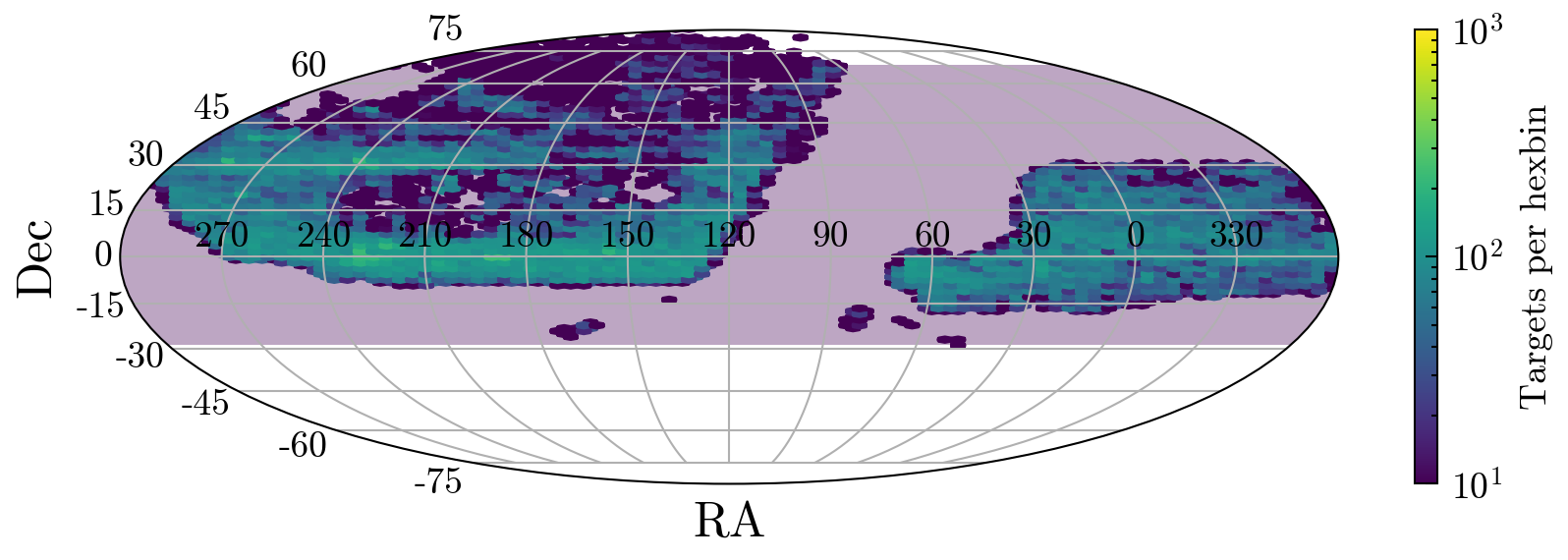} 
\caption{The sky coverage for the DESI quasar C~{\sc iv} RM sample used in this feasibility study (refer to the sample description in Section \ref{sec:sample}). The colour map represents the density of targets. The ZTF sky area for the \textit{r}-band is shown in the purple block-shaped region with a declination range of $-28$\textdegree $<$ dec $<66$\textdegree. Only 1\% of the quasar sample lies outside ZTF's sky coverage.}
\label{fig:sky}
\end{figure*}

\section{Mocks} \label{sec:mocks_method}

A mock pipeline builds confidence in the developed method by demonstrating that it recovers the expected results when applied to simulated data for which we know the true time lag. Furthermore, it allows one to determine the accuracy of stacked RM for different numbers of quasars with varying data quality. Other RM studies have conducted similar simulations to test their methods and we have inherited many of the techniques from them as presented below \citep{King2015, Yu2020, FonsecaAlvarez2020, Malik2022, Czerny2023}.
In the following subsections, we describe the selection process and characterisation of the C~{\sc iv} DESI quasars sample, how we bin them for stacking, and a description of the mock light curve generation for the quasars in each bin.

\subsection{Survey Assumptions for Forecast and Sample Description} \label{sec:sample}


DESI is primarily designed to maximise the number of galaxies and quasars whose redshifts can be measured for the advancement of key cosmological objectives. Revisiting of targets on the sky is typically not prioritised unless it is to obtain a greater spectral SNR. 
The DESI main survey aims to undertake nine passes across the footprint enabling all 3.4 million quasar targets to be observed over the lifetime of the eight year survey \citep{DESI2022}. The geometry of the robotic fibre positioners limits the number of targets observed in a particular field. Therefore, DESI will pass over the same area of sky multiple times. Each time the objects observed within a field is changed in order to fill in the gaps and increase coverage. Quasars that lie in the overlapping areas could potentially be observed multiple times. Of such targets, 0.8 million Lyman-alpha quasars at $z > 2.1$ will, by design, be reobserved at least four times \citep{Schlafly2024} to improve the SNR of the Ly$\alpha$ forest absorption. These repeat observations in this redshift range serendipitously benefits C~{\sc iv} RM studies. Additionally, due to weather and observation conditions, objects not specifically identified for multiple passes may end up with exposures split over multiple days. Figure \ref{fig:sky} shows the DESI footprint mapped out by the EDR and DR1 observations; several regions of the sky have already incurred multiple passes building up significant data. In this section we describe the use of the DESI survey data and survey strategy to build mock data with which we can test the robustness of the stacked RM technique.

In order to characterise the quasars we would use for a DESI RM program, we define a sample by applying cuts to all targets that are spectroscopically classed as quasars from DR1. For the C~{\sc iv} line to appear in the DESI spectrum, along with its blue wing and some continuum for continuum subtraction, a redshift cut of $z$ > 1.48 is applied. 

Photometric band flux can have a significant contribution from broad emission lines (e.g., \citealt{Mckinney2023, PozoNunez2023}). By examining the DESI spectra, we determined that strong C~{\sc iv} emission lines can contribute up to 30\% of the ZTF band flux, with a median contribution of 13\% for high SNR lines. Since we propose to use photometry data to construct continuum light curves, we ought to choose the photometric band carefully in order to avoid broad line contamination that will lead to aliases or a null time lag detection \citep{Czerny2023}. Figure \ref{fig:filters} illustrates an example of where the most prominent broad lines, C~{\sc iv} and Mg {\sc ii}, lie in relation to the ZTF filters for a quasar at redshift 2. The \textit{r}-band evades both C~{\sc iv} and Mg {\sc ii} broad lines from falling into its wavelength range and so for this example we would obtain the \textit{r}-band light curve to represent the continuum. In a similar fashion, the photometric band suitable for each quasar is a function of its redshift. We picked the band based on redshift ranges in which neither C~{\sc iv} and Mg {\sc ii} of a typical width contaminate the band, and when more than one band is available we prioritise bands of high cadence in the order of \textit{r}, \textit{g}, then \textit{i}, giving the following redshift ranges:
\begin{equation}
\begin{split}
    1.67 \leq z < 2.48 : \text{\textit{r}-band}, \\
    2.48 \leq z < 2.72 : \text{\textit{i}-band}, \\
    2.72 \leq z < 3.87 : \text{\textit{g}-band}, \\
    z \geq 3.87 : \text{\textit{r}-band}. 
\end{split}
\end{equation}
Between redshifts 1.48 and 1.67 none of the bands are contaminant free so this forces us to raise the redshift cut of the sample up to $z \geq 1.67$. Additionally, a magnitude cut is applied to reflect the ZTF \textit{r}-band magnitude limit of 20.6 (throughout this paper we use the AB magnitude system) since the majority of the sample will have \textit{r}-band light curves. Figure \ref{fig:lumzdistbins} illustrates these cuts as they appear on the luminosity-redshift plane of the quasar sample.

To ensure broad line detection and accurate flux measurements, we apply a SNR > 15 cut for the C~{\sc iv} line calculated using values from the FastSpecFit VAC (Moustakas et al. in prep). We kept the SNR cut relatively high as the values produced by FastSpecFit are calculated from coadded spectra so the SNR calculated is an overestimation of the C~{\sc iv} SNR of individual epochs. This SNR cut retained 86\% of the sample, with an average C~{\sc iv} flux SNR of 76. If a larger sample is required for the stacking procedure this cut may be lowered. Finally, we limit the sample to quasars with $\geq 2$ observations on separate nights. DESI cumulates spectra of objects each night through the combination of shorter exposures. Any repeat observation taken on the same night would be coadded for higher SNR. The final sample has 228,937 quasars spanning a redshift range of 1.48 < $z$ < 5.2 and a luminosity at 1350\AA{} range of $ 44.72 \leq \log \lambda L_{1350 \AA{}} / \mathrm{erg\,s^{-1}} \leq 47.10 $.


\begin{figure} 
\includegraphics[width=8cm]{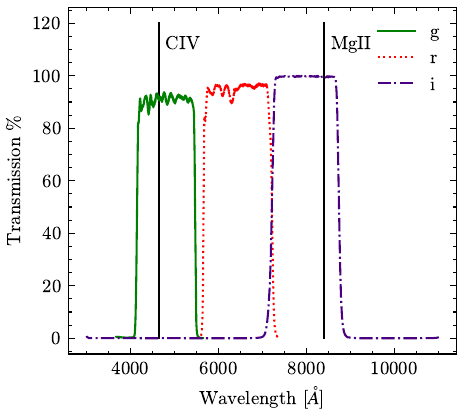} 
\caption{The filter transmission functions for the $g$, $r$, and $i$ photometric bands used in the ZTF survey. The solid black lines provide an example of the central locations of the observed frame C~{\sc iv} and Mg {\sc ii} emission lines for an object at a redshift of 2. In this case, it is important we avoid g and i bands which are contaminated by C~{\sc iv} and Mg {\sc ii} broad lines respectively. Rather, we pick the \textit{r}-band to ensure the flux for the continuum light curve is not affected by bright broad emission lines.}
\label{fig:filters}
\end{figure}

\begin{figure} 
\includegraphics[width=8cm]{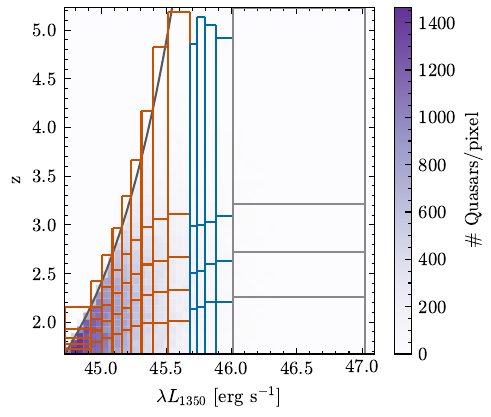}
\caption{The luminosity-redshift distribution of the C~{\sc iv} RM sample is displayed by the purple colour map and is overlaid with the luminosity-redshift bins. The samples redshift range is 1.67 < $z$ < 5.23. The redshift ($z$) lower limit at $z$ = 1.67 is set by requiring the presence of the C~{\sc iv} broad line in the DESI spectrum and having a photometric band available without broad line contamination. The range of the rest frame 1350\AA{} luminosity of the sample is $ 44.72 \leq \log \lambda L_{1350 \AA{}} / \mathrm{erg\,s^{-1}} \leq 47.10 $. The \textit{r}-band magnitude limit at 20.6 mag (imposed by the ZTF \textit{r}-band sensitivity limit) is shown as the dark grey limiting curve; the brightest \textit{r}-band magnitude in the sample is 15.2. 10 equal-population bins are created in luminosity but due to the large width of of the highest luminosity bin, shown in grey, it is discarded. The next highest luminosity bin is further divided into four bins, highlighted in blue. Then five equal-population bins in redshift are created per luminosity bin, with 3687 quasars per luminosity-redshift bin (in orange), except for the highest luminosity bins which are only divided into four redshift bins to maintain a significant sample of 921 quasars per luminosity-redshift bin (in blue).}
\label{fig:lumzdistbins}
\end{figure}

\subsection{Luminosity-Redshift Binning}
\label{sec:binning}

Binning in luminosity and redshift are chosen as they are factors directly present in the $R$-$L$ relation we are using. In reality, the size of the BLR is not solely dependent on the luminosity but, to a lesser extent, a range of other factors such as the accretion rate \citep{MartinezAldama2019}. The accretion rate, quantified either by the Eddington ratio or its equivalent, the dimensionless accretion rate, has been shown to have a significant influence on the BLR \citep{Wang2013, Du2016, Panda2023}. Further binning in accretion rate to reduce lag scatter may be ideal, especially if using an $R$-$L$ relation where the accretion rate is a variable. Even so, the number of variables binned by and the number of bins created are limited by the sample size. On one hand, the final lag error is reduced by creating a bin wide enough to have a substantial number of quasars for stacking. On the other hand, narrower bins will reduce the diversity of AGN, and hence BLR radii, meaning that we stack a narrower range of expected time lags and thus reduce the error in the final stacked lag. It is important that the right balance must be achieved.

Bearing this in mind, the sample of quasars is divided into equal-population bins such that there are an equal number of quasars per bin for the purposes of stacking. The bins are created to ensure that the quasars within them are of similar nature and will have similar sized BLRs. We use the distribution of monochromatic luminosities at 1350\AA{}, typically used for C~{\sc iv} RM, to create ten bins of luminosity. These luminosities are derived from the $r$-band magnitudes as follows: we compute the synthetic $r$-band magnitude of the \citet{VandenBerk2001} quasar composite spectrum, normalised to an assumed bolometric luminosity  of $10^{46}\mathrm{erg\,s^{-1}}$, and integrated over the ZTF $r$-band filter transmission curve. This provides the conversion factor from the observed $r$-band magnitude to monochromatic luminosity at 1350\AA{}.

DESI's vast quasar sample enables us to further divide each luminosity bin into five equal-population bins of redshift, resulting in 3687 quasars per luminosity-redshift bin. This redshift binning serves two purposes. First, since the $R$-$L$ relation is not known to evolve cosmologically \citep{Khadka2021} and is not part of the model for the mock light curve simulations, any redshift dependence we detect would indicate systematic errors in the method. Second, when applying this method on data in future studies, redshift binning allows us to test for potential redshift evolution of the quasar population.

Since we use an equal-population binning approach, the high redshift-high luminosity bins are inevitably wider as fewer quasars are observed in these ranges. As such, these bins are more susceptible to poor time lag measurement. The high luminosity-high redshift quasar space has a lower density of observations due to limitations in telescope diameter and observing time. High luminosity sources are also rarer so require observation of a larger cosmic volume (i.e., a larger areal coverage of the survey) to access sizeable populations. This results in the highest luminosity bin being too wide for the purposes of stacked RM since we can no longer assume the quasars in this bin have similar lags. In order to restrict the range of lags, we further divide the highest luminosity bin into five luminosity bins with four redshift bins each to maintain a significant sample of 921 quasars per luminosity-redshift bin. Within this further binning, we still discard the highest luminosity bin for two reasons. First, this bin is still too wide to assume the quasars within it will have similar lags. Secondly, the quasars in these bins may have very high observed lags (>1000 days) due to a combination of large BLR sizes and time dilation. Plus, high luminosity quasars exhibit less variability which weakens our ability to measure an accurate time lag \citep{Wilhite2008, Meusinger2011}. Paired with a high redshift, the variability is further smoothed out by time dilation. These factors means that, for the lags of these quasars to be measured, it is especially necessary to have spectroscopic observations spread out further out in time and extended over a longer period \citep{Kaspi2021}. However, we are limited by the DESI survey design and baseline to probe this extremity of the quasar population space.

The bins are presented in Figure \ref{fig:lumzdistbins}. In total, there are 61 luminosity-redshift bins (9 luminosity $\times$ 5 redshift + 4 luminosity $\times$4 redshift) and each produces one point on the $R$-$L$ plot. This binning is suitable since it guarantees that the lags within each bin are sufficiently similar for stacked RM to work as well as being well-populated. We test if the bin size can be further optimised in Section~\ref{sec:binSize}.

Binning the sample in the manner described gives us at least $\sim$1000 quasars in each bin. It is impressive that DESI DR1 can already provide so many quasars for each bin. Data from subsequent years of observation will allow for more prudent choice of quasars (e.g. those with higher SNR or more spectral epochs). We test how the number of quasars stacked per bin affects the recoverability of the time lag in Section \ref{sec:numStack}.

\subsection{Light curve simulation} \label{sec:lcsim}

We model the variations with a light curve simulation code developed for the OzDES-RM campaign as described in \citet{Penton2022}. For the main run, 250 mock light curve pairs are simulated per bin. We stack 250 quasars as a benchmark but this is only a fraction of the thousands of DESI quasars available per bin (see Figure \ref{fig:lumzdistbins}), we explore stacking different numbers of quasars in Section \ref{sec:numStack}. The light curve simulation takes into account the intrinsic properties of quasars in each bin as well as the observational characteristics of the DESI survey for the C~{\sc iv} light curve (described in Section \ref{sec:sample}) and the ZTF survey for the continuum light curve.

There are two steps in simulating the light curves. Firstly, we need to generate the underlying continuum and emission line light curves. Secondly, the underlying light curves are sampled to produce the forecasted observed light curves.

In the first step, the underlying continuum light curve is simulated with a DRW model based on the specified intrinsic properties of a quasar. A DRW is a random walk which has a restoring term that constrains this variability about a mean value, imposing a characteristic timescale on the variations. The theory that the observed continuum variability of a quasar is a stochastic process satisfactorily described by a DRW model is substantiated by most data currently available \citep{Kelly2009, Macleod2012, Fagin2023}. It is one of the simplest statistical processes which successfully mimics a stochastically varying quasar light curve that habitually saturates at some amplitude \citep{Stone2022}. Quasar variability is determined by factors such as luminosity and rest-frame wavelength. We provide these by characterising each mock object by a randomly drawn luminosity and redshift within its bin's bounds. We verified that the simulated variability matches the continuum variability of ZTF light curves, for further details see Section \ref{sec:limit_obs}. The claim that the DRW model satisfactorily describes stochastic quasar variability and the contended exceptions to it are discussed further in Section \ref{sec:discussion}. 

The emission line light curves are then assumed to be the lagged, smoothed, and scaled versions of the continuum light curves. The assigned luminosity to the mock quasar is also used to calculate the input `observed' time lag using the $R$-$L$ relation from \citealt{Hoormann2019},
\begin{equation}
\log R \text{[days]} =  0.82 + 0.49 \log\frac{\lambda L_\lambda(1350\text{\AA})}{10^{44}[\mathrm{erg\,s^{-1}}]},
\label{eqn:HoormannRL}
\end{equation}
and shifting it to the observed frame, where $\lambda$ is the wavelength set to 1350\AA{}. This literature  C~{\sc iv} $R$-$L$ relation is fitted from compiled C~{\sc iv} data sets spanning a luminosity range approximately between $10^{39} \mathrm{erg\,s^{-1}}$ and $10^{48} \mathrm{erg\,s^{-1}}$ encompassing the luminosity range we are probing. We will be using the average lag per luminosity-redshift bin as the target lag to recover in that bin. The standard deviation of the simulated lags, calculated with equation \ref{eqn:HoormannRL}, is $\lesssim$ 10\% of the average lag in each luminosity-redshift bin. We use this relation without adding any dispersion in order to focus on the pipeline's ability to recover lags from sparse spectroscopic data. In reality, RM observations of C~{\sc iv} lags display a large scatter that is up to $\sim$10 times larger than what we have simulated for a given luminosity bin \citep{Shen2024}. However, introducing this scatter into the mocks will prevent us from decoupling the scatter induced by the pipeline from the intrinsic scatter of the $R$-$L$ relation. Testing the mock pipeline with quasars that deviate from the $R$-$L$ relation is left to future work.

The mock underlying light curves are generated across six years to cover the first five years of the DESI survey plus a year preceding it in order to obtain a maximal photometric window and to have continuum data prior to the spectroscopic observations. The longer the continuum light curve the better the continuum variations can be characterised before identifying the lag in the emission line variations. The emission line light curve will be generated from the continuum using a top-hat function with width proportional to the time lag (for more information see Section \ref{sec:lag_measurement}). 

The suite of underlying light curves are stored and used for all the different runs of the pipeline described in Section \ref{sec:results}. In other words, we have kept the sample of mock quasars constant throughout the various tests.

The second stage of light curve simulation is to sample the underlying light curves to simulate the observed light curves. Whereas the underlying light curves or the mock quasar sample is kept constant in the different runs, the way we `observe' or sample them is varied. The continuum light curve is sampled according to ZTF's cadence which is nominally 2-3 days. However, for light curves spanning a few years, seasonal gaps appear in the light curves. Plus, other smaller gaps appear due to weather, moon interference, and downtime. To simulate a realistic photometric cadence, we analysed ZTF light curves of DESI counterparts from our sample and created distributions of cadence and seasonal gap length to draw from in the sampling of the continuum light curves.

For the emission line light curves, the DESI observations of a particular target are predicted from the DR2 data distributions of the baselines (number of days between first and last observation), the number of total repeat observations, and their cadence. Figure~\ref{fig:cadence} shows the distributions of baselines and cadences from DR2 used for our mock light curve sampling. The median number of spectroscopic epochs, observed on separate nights, is three days, the minimum, as defined by our sample cut, is two days, and the maximum is 13 days. The median cadence is 15 days. These dates are picked within a 5-year baseline and then the photometric light curve dates are picked to span 1400 days prior to the last spectroscopic observation, a baseline that reflects the current data available. Figure \ref{fig:examplecadence} showcases the simulated spectroscopic and photometric cadence for 1000 quasars from the C~{\sc iv} sample. The typical set of expected DESI repeat observations occur in short bursts and will occasionally have extra observations in the following years during the observational seasons. The distribution of the time separation between spectroscopic-photometric light curve pairs exhibits trends that derive from the seasonal window function. This data represents mock light curves of DESI DR2 data - repeat observations that occur during the rest of the survey beyond DR2 data will extend our current baseline range and increase the number of repeat observations for a subset of these targets. We test the improvement in time lag recovery with the addition of extra observations in Section \ref{sec:extra}. 

\begin{figure} 
\includegraphics[width=8cm]{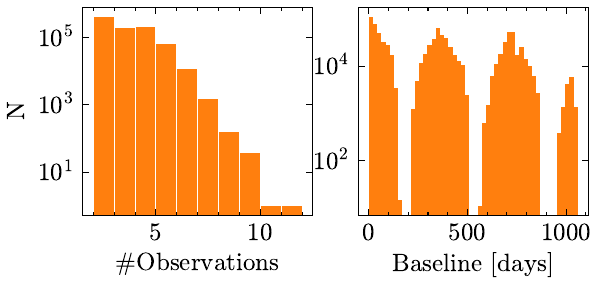}
\caption{Left: Distribution of the number of repeat observations on separate nights per target ($\geq$2). Right: Distribution of the baseline of observations per target i.e. number of days between first observation and most recent observation. This spans the first three years of DESI data and we can start to see the observational seasons in the DESI survey. The targets observed within the first year are revisited in subsequent years.}
\label{fig:cadence}
\end{figure}


\begin{figure*} 
\includegraphics[width=\textwidth]{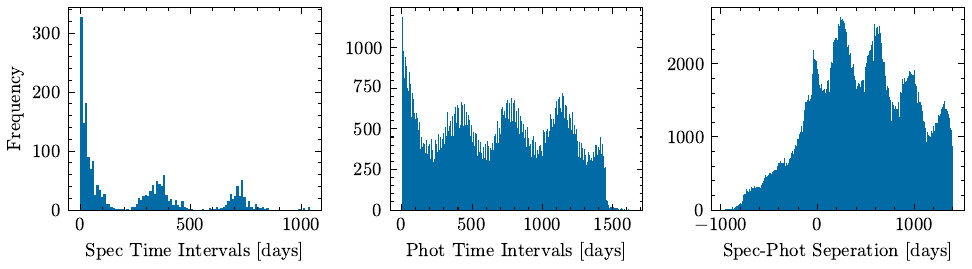}
\caption{A set of plots used to illustrate the spectroscopic and photometric cadence of the mock light curve suite. These plots have been generated from the cadences of a representative sample of 1000 mock light curve pairs. Left: A histogram of the time interval between the first and subsequent epochs of the mock spectroscopic light curves. Middle: A histogram of the time interval between the first and subsequent epochs of the mock photometric light curves. Right: Time separation of all spectroscopic-photometric epoch pairs.}
\label{fig:examplecadence}
\end{figure*}

The last step is to add noise to each of the sampled light curves. The noise is calculated for each epoch as $\sigma X$ where $\sigma$ is drawn from the error distribution of the corresponding data set and X is randomly picked from the set \{-1, 0, 1\}. Validation tests comparing this error-generation approach with Gaussian-distributed errors, which includes values both below and above 1$\sigma$, showed no material difference in the resulting lag recovery. Figure \ref{fig:errs} shows the distributions of ZTF and C~{\sc iv} fractional flux errors used to draw $\sigma$ from for the continuum and C~{\sc iv} light curves respectively. The ZTF error distribution is compiled from ZTF light curves of DESI counterparts, converted from \textit{r}-band magnitude to fluxes. The C~{\sc iv} flux errors are the errors of boxcar integrated fluxes of DESI C~{\sc iv} line profiles after continuum subtraction and accounting for Milky Way transmission over a wavelength range $\lambda_{\text{obs}}-n_{\sigma} \sigma <\lambda< \lambda_{\text{obs}}+n_{\sigma} \sigma$ where $\lambda_{\text{obs}} = \lambda_{\text{rest}}(1+z)$ with $\lambda_{\text{rest}}= 1549$\AA{} for C~{\sc iv} and $z$ is the quasars redshift, $n_{\sigma} = 3$, and $\sigma$ [\AA{}] $= \sigma_{\text{fsf}}\lambda_{\text{obs}}/c $ where $\sigma_{\text{fsf}}$ is the Gaussian emission line width calculated by FastSpecFit in km/s and $c$ is the speed of light.  We place upper limits on the error distributions such as to remove outliers. We do not simulate outliers as when working with data the light curves would be processed to remove outliers or artefacts of any kind.
The same error distributions are drawn from again to assign an error to each light curve point, such that the noise and errors of each light curve are not correlated. There is also no correlation between the errors in light curve pairs as the continuum and emission line data are from completely different sources.
\begin{figure} 
\includegraphics[width=8cm]{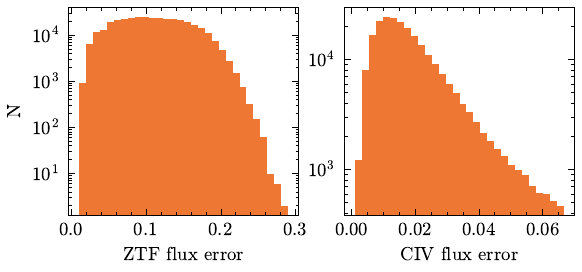}
\caption{Distribution of ZTF (left) and C~{\sc iv} (right) fractional flux errors used to simulate the errors in the mock photometric and spectroscopic light curves respectively. Both distributions are fractional flux errors in units of $10^{-17}$ erg cm$^2$ s$^{-1}$.}
\label{fig:errs}
\end{figure}

Figure \ref{fig:lcs} presents an example of mock photometric and spectroscopic light curves generated as described. The quasar simulated here has a redshift of 1.63 and a luminosity at 1350\AA{} of 5.7$\times 10^{44}\mathrm{erg\,s^{-1}}$. Using Equation \ref{eqn:HoormannRL}, the observed lag between the two light curves is 40.6 days. The solid lines are the simulated underlying continuum and lagged emission line light curves. The points show the sampling of how each of these might be observed. The units and amplitude of flux in our simulated light curves are arbitrary. RM is not purportedly sensitive to the absolute scale or scale differences between the two light curves, what matters is the shape of the light curve. We tested how varying the order of magnitude of the fluxes of both light curves effects the lag measurement and we found it to be stable. When producing light curve pairs from data we convert the photometric light curve in units of magnitude to units of flux such that they both have the same units.

\begin{figure*} 
\includegraphics[width=\textwidth]{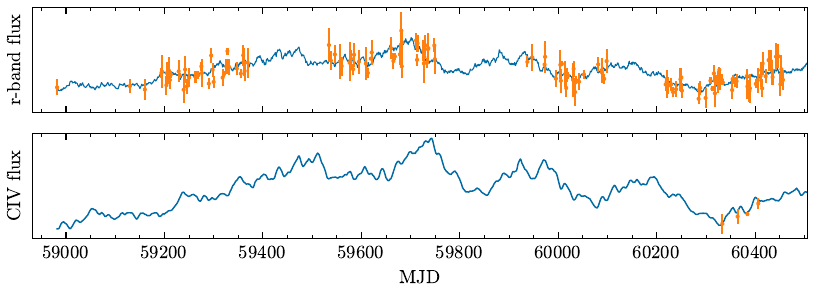}
\caption{Example mock light curve pair. The solid blue lines are the underlying simulated light curves for the continuum and C~{\sc iv} emission line flux. The emission line light curve is the scaled and lagged continuum light curve. The orange points show how the observed light curves are sampled from the underlying light curve for ZTF photometry and DESI spectroscopy in the continuum and C~{\sc iv}  light curves respectively. This is simulated for a quasar at redshift 1.63 and with luminosity at 1350\AA{} of 5.7$\times 10^{44} \mathrm{erg\,s^{-1}}$. It has an observed lag of 40.6 days.}
\label{fig:lcs}
\end{figure*}

As part of the process of checking the similitude of the simulated light curves to the data, we recreate all the various input distributions from the simulated light curves and directly compare the shape and amplitude of each to the data distributions. These include distributions for the quasar sample such as those of redshift, luminosity, and variability. We also reproduced distributions of the flux amplitude, flux errors, cadences, baselines, number of observations, and the seasonal gaps for both the photometric and spectroscopic light curves. For all checks, the simulated light curves matched the data well.

\section{Lag Recovery Methodology} \label{sec:lag_measurement}
A wide range of time lag analysis methods have been developed and applied in past RM campaigns. The standard approach is the cross-correlation function (CCF), along with several variations such as the interpolated cross-correlation function (ICCF; \citealt{Gaskell1987, White1994}), the discrete correlation function (DCF; \citealt{Edelson1988}), and the z-transformed discrete correlation function (zDCF; \citealt{Alexander2013}). However, correlation-based methods do not leverage our understanding of the stochastic nature of quasar variability and are sensitive to sparse sampling and time binning in correlation space.
To address these limitations, alternative techniques have been developed, including the von Neumann estimator \citep{Chelouche2017}, Gaussian process regression \citep{Wilkins2019}, Bayesian inference–based codes such as {\sc litmus} \citep{Mcdougall2025} and others which utilise Markov Chain Monte Carlo (MCMC)  strategies such as {\sc javelin} \citep{Zu2011}, {\sc cream} \citep{Starkey2016}, {\sc pyroa} \citep{Donnan2021}, and {\sc mica} \citep{Li2013}.

While correlation functions- especially ICCF- are considered more traditional, they are poorly suited for DESI light curves, which are unevenly sampled and sparsely populated. Interpolating between such incomplete spectroscopic data points may yield unreliable results, as the interpolated light curves may not adequately capture the true underlying variability. The zDCF is a potential alternative, as it is designed for irregular and sparsely sampled time series, does not assume a specific AGN variability model, and naturally lends itself towards stacking procedures \citep{Alexander2013}. However, it is ultimately unsuitable for DESI light curves due to its requirement of at least 12 observational epochs.

A parametric approach implemented using MCMC, a popular choice being {\sc javelin} (Just Another Vehicle for Estimating Lags In Nuclei; \citealt{Zu2011}), is more suitable for the spectroscopic data we have available. We explain this in further detail in the next section. Therefore, after producing the mock continuum and emission line light curves for each object, we have chosen to compute the time lag between them using {\sc javelin}. This is what we have defined as the main pipeline. We have also tested an alternative pipeline using CCFs instead, this is presented in Section \ref{sec:CCF}.

\subsection{JAVELIN}
\label{sec:javelin}
{\sc javelin} models the continuum light curve variability using a DRW and models the emission-line response with a parametrised top-hat transfer function (Equation \ref{eqn:transfer}). It employs a Bayesian framework with a two-step MCMC algorithm to estimate time lags \citep{Zu2011}. The continuum is known as the driving light curve and in the first step the continuum variability is constrained by a DRW model to compute the posterior probability density distributions for two parameters: 1) the variability amplitude of the continuum, $\sigma$, and 2) the characteristic damping timescale, $\tau_{\text{damp}}$. Using the previous two distributions as priors, the second step maximizes the likelihood by simultaneously fitting a DRW model to the two light curves with three more parameters. The continuum light curve is interpolated, shifted, smoothed and scaled assuming a top-hat smoothing transfer function centred at 1) $\tau$, the time lag, with 2) full width $w$ and 3) flux scaling factor $s$ between the line and continuum amplitudes A\textsubscript{line}/A\textsubscript{cont} to fit the emission line light curve using {\sc javelin}'s spectroscopic Rmap module. The time lag posterior is obtained by marginalising over all other parameters.

The ability to use Bayesian inference techniques when modelling the variability in parametric approaches gives the advantage of better qualification of uncertainties in the lag measurement. There have been studies comparing these methods using simulations. \citet{Li2019} compared ICCF, zDCF and {\sc javelin}. They found {\sc javelin} outperformed the other methods across many assessment criteria and did especially well with limited light curve quality. \citet{Yu2020} found {\sc javelin} provides better error estimates in comparison to ICCF. Furthermore, many observational RM studies have used multiple time lag methods to check for consistency. Examples include \citet{Shen2015}, \citet{Zajacek2019}, and \citet{Homayouni2020} who all found {\sc javelin} produces consistent lags with the other methods. Based on this validation, we adopt {\sc javelin} as the most suitable method for extracting time lags, given the characteristics of our dataset. Some of our light curves contain as few as two epochs, rendering many non-parametric time lag estimation methods ineffective due to their stronger dependence on data density. In contrast, {\sc javelin}'s Bayesian MCMC framework allows for the exploration of a wide range of model parameters, meaning the interpolation performed by {\sc javelin} is grounded in empirical modelling of quasar variability and does not rely on densely sampled light curves. A recent study explicated the extent to which MCMC-based lag estimation techniques substantially outperform CCF approaches, particularly in addressing the challenge of seasonal aliasing \citep{Mcdougall2025}. For these reasons, we initially develop the stacked RM using {\sc javelin}. For completion and comparison with previous studies, we also redevelop a stacked RM pipeline with CCFs which is later presented in Section \ref{sec:CCF}. 

None of the model parameters are fixed in the light curve simulations just as they would not be in nature. However, the limited DESI cadence is not sufficient to constrain the damping timescale and the top-hat function width. For example, \citet{Zhou2024} and \citet{Kozlowski2016} have emphasised that long baselines are needed to constrain the damping timescale. Leaving these as free parameters result in a unphysical transfer function fit; so, we fix them when measuring the time lag, as done in previous studies such as \citet{Homayouni2020}. Doing so has a negligible effect on the lag recovery and has the advantage of producing cleaner lag distributions. In this project, the damping timescale is fixed to 700 days. We have set the damping timescale to the median value in the simulated light curves and, whilst the damping timescale typically ranges from 200-1000 days \citep{Kelly2009, Macleod2012}, this value coincides with the value given in one of the longest timescale variability studies by \citet{Stone2022} who found a median of 750 days. By setting a fixed damping timescale, the first step of {\sc javelin} is reduced to "interpolating" the continuum light curve using a DRW \citep{Li2019}. Due to the sparse data in the spectroscopic light curve, we do not explicitly fix the damping scale in the second step, as {\sc javelin} has difficulty converging under this constraint. Instead, we confine the damping scale to a very narrow range between 699 days and 706 days which evades the numerical MCMC issues.

Similarly, in order to avoid analogous issues we came across at long chain lengths, rather than fixing the top-hat width, we have set a flat prior range of 3 < $w$ < 40 days instead. This covers the entire range of $w$ in the simulations, in which the transfer function width is set to 0.1$\tau$ for the smoothing timescale to be sufficiently shorter than the time lag \citep{King2015, Penton2022}. It has been previously shown that the lag measurement is insensitive to setting the damping timescale and top-hat width to different values (e.g., \citealt{Li2019}, \citealt{Yu2020}). We have verified this degeneracy on a small sample of our simulations too and the exact choice should not effect our final results.

Finally, we define a flat prior for the time lag. We picked it to be restricted between 0 and 500 days in the observed frame. Since we relinquished the highest luminosity bin, the highest luminosity  and redshift combination possible in the remaining bins ($1\times 10^{46}\mathrm{erg\,s^{-1}}$, 5.23) will at most generate a lag of $\lesssim 400$ days. Therefore, our lag prior covers the entire range of simulated lags; across the suite of simulated light curves, the minimum observed lag is 33 days and the maximum is 383 days. We explore how the choice of time lag prior effects the recovered lag in Appendix \ref{app:prior}.
{\sc javelin} has a feature that logarithmically penalizes lags larger than a fraction of the continuum baseline; we set this to one to ensure no lags are penalised. 

For each pair of objects we use 50 walkers and 1000 burn-in steps. We run the chain until it has converged. That is, when the autocorrelation time for all parameters changes less than 1\% and the number of MCMC iterations is larger than 100 times the autocorrelation estimate \citep{Foreman-Mackey2013, Read2020}. We set the maximum number of chain steps to 5000 but it rarely takes this long to converge. In general, light curves that have fewer epochs and are sampled at less optimal intervals require more time to converge.

The variation in the relatively short spectroscopic light curve may match with several sections of the photometric light curve. Paired with the finite sparse nature of the light curves, {\sc javelin} almost always produces a noisy multi-modal posterior distribution. With real data, false correlations between the two light curves may appear from a combination of residual continuum remaining in the emission line even after subtraction or some broad line emission penetrating into the continuum photometric band. It is our aim to discern whether the correct time lag peak will arise from amidst the aliases and noise during the stacking procedure.

\subsection{Stacking lag distributions}
Stacked RM was introduced by \citet{Fine2012} to overcome obstacles in traditional RM by stacking quasar samples which have well-sampled continuum light curves and only a few, perhaps as few as two, emission line measurements. The viability of this method has been demonstrated in previous studies by stacking CCFs (\citealt{Fine2013}; Eltvedt et al. in prep), ICCFs \citep{Malik2024}, and zDCFs \citep{Li2018}. Instead, this paper presents a new approach of stacking MCMC-derived posterior distributions of the time lag in the observed frame of reference. In our approach we are making the assumption that our luminosity-redshift bins are thin enough such that we are stacking objects with very similar BLR radii as justified in Section \ref{sec:binning}.

The individual time lag distributions are additively stacked, which we argue below is more suitable for our data than alternative stacking approaches. Radial stacking, in which the prior for subsequent lag measurements is updated using the previous posterior distribution, is not suitable for sparse light curve data, as a single object's posterior may be erroneously biased towards a false peak or be too noisy to provide useful information. Another way is hierarchical stacking which maximises the information derived from the individual posteriors using additional hyperparameters \citep{Brewer2014}. However, this adds an extra level of complexity and computational expense which is impractical for large samples unless it is further optimised by computational techniques such as importance sampling. Multiplying the posterior probability distributions would only be valid under the assumption that the priors are independent. In our case, this assumption is violated because binning the quasars by similar lags induces dependencies among the priors of the light curve parameters. We have applied an additive stacking procedure by appending the MCMC time lag chains for all the objects in each bin. Admittedly, this is not a mathematically robust way of combining posterior distributions. Nonetheless, this approach achieves the goal of enhancing the genuine lag signal by elevating the portion of the posterior distributions that is consistently more concentrated. In any case, we assess the robustness of the outcomes obtained with this method by comparing them to the known correct result from the simulations.

After stacking, a clear time lag peak emerges from the once apparently noise-ridden distributions. For clarity, we designate the term `peak' to mean the distribution surrounding the maximum. The peak is identified and distinguished from the rest of the distribution by first convolving the discrete stacked posterior distribution with a Gaussian filter with a dispersion of one day to derive a smoothed distribution (a similar method is used in \citealt{Grier2019}). We isolate the peak by determining the lag interval it occupies and discard the rest of the distribution. This interval is found from the relative minima and points of inflection in the distribution. The extrema on either side of the maximum are located. We make a check ensuring that the distance of the peak to each end of the interval is greater than or equal to a scale which we set to 20 days. This is in order to account for doubly peaked lags or points of inflections near the peak that will result in an underestimation of the lag error. If the interval is too tight on either end, we expand it by shifting the edge to the next extrema on that side. The same check is conducted recursively until satisfied. If the peak lag occurs before or after all the extrema then the edge of the interval is set to the corresponding edge of the whole distribution. Qualitatively, this amounts to bounding the maximum of the smoothed distribution by the minima that lie on either side. This methodology is motivated by the approach of \citet{Grier2019}, which was employed to mitigate aliasing effects in the distribution.

The final lag measurement is defined as the lag corresponding to the maximum of the peak, referred to here as the maximum a posteriori (MAP) estimate. The associated uncertainties are derived from the 16th and 84th percentiles of the peak distribution. 
We compared the use of the MAP versus the median (50th percentile) of the peak as the lag estimate. We found that the MAP generally yielded more accurate recovery of the input lags. The MAP had an average deviation of 13\% from the mock lag, while the median deviated slightly further, on average by 15\%.
Due to the asymmetric nature of some peak distributions, using the MAP instead of the median occasionally leads to unphysical uncertainty intervals, where the lower error exceeds the lag value (i.e., the lag is below the 16th percentile) or the upper error is smaller than the lag (i.e., the lag exceeds the 84th percentile). In such cases, the affected error is redefined to be symmetric with the opposite error with respect to the lag.

We also considered the possibility of null lags arising from the MCMC sampling procedure. We are particularly aware of the issue of walkers becoming stuck at prior boundaries. The probability of this artefact appearing increases with sparse light curves that generate broad posteriors and can lead to artificial peaks near the lower lag limit of zero days. In such rare cases, where only a partial (half) peak is present, symmetric errors are again imposed. Furthermore, to avoid including unreliable lag measurements, we applied a quality cut that excluded any quasars with recovered lags less than five days in the final $R$-$L$ fits.
It is important to note that these corrective measures are not required in the main mock run, They are only triggered in specific robustness tests (Section~\ref{sec:results}) where the light curves are intentionally degraded, and even then occurred in no more than a few (<3) bins.

Figure \ref{fig:stackedpd} shows an example stacked time lag distribution with the input and recovered lag for that luminosity-redshift bin marked. After the stacking procedure, the noisy multi-modal individual lag distributions build up to consistently reveal a clear lag peak that aligns with the input lag. The width of the lag peak is only partially accounted for by the diversity of lags in the stack. In fact, there is no correlation found between the width of the lag peak and the dispersion or range of simulated lags in the bin. The shape of the peak is controlled by observational factors such as cadence.
A quality assessment of the lag measurement is made by calculating the SNR of the lag peak defined as the peak-over-noise ratio, i.e., the maximum of the distribution over the average value of the distribution outside the peak region. The SNR values in this study may seem small ($\gtrsim 1$) but values greater than 1 in the context of our stacking method signify a successful build-up of the lag signal.

\begin{figure}
\includegraphics[width=8cm]{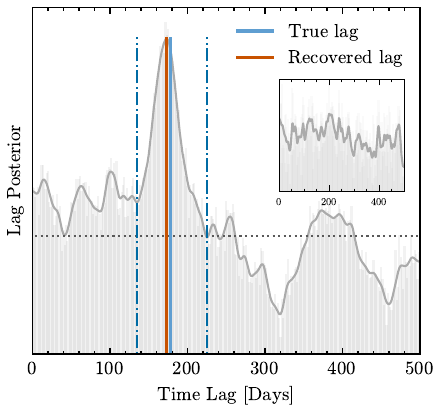}
\caption{Example of a stacked time lag posterior distribution consisting of 250 quasars in a luminosity-redshift bin with an average simulated time lag of 178 days (solid light-blue line). The y-axis in this and other lag posterior plots is shown on a logarithmic scale to make the shape of the posteriors easier to see. The objects in this bin had an average luminosity at 1350\AA{} of 5.7x10$^{45} \mathrm{erg\,s^{-1}}$ and an average redshift 2.7. The orange line is the recovered lag of 173 days. The {\sc javelin} MCMC walkers probed a 0-500 days time lag range. The embedded subplot on the right shows the lag posterior distribution of just one of the quasars that contributed to the stack -  it is noisy and multi-modal. Although individual lag distributions do not give a reliable lag peak, it can be seen a clear peak emerges after stacking and this is identified using a Gaussian filter (solid grey line). The bounds for the peak used in determining the lag uncertainty and SNR are shown as dash-dotted dark-blue lines. The horizontal dark grey dotted line represent a SNR value of one to separate noise from significant peaks.}
\label{fig:stackedpd}
\end{figure}

A summary of the whole pipeline including both the simulation of mock data from actual data and the lag recovery method is presented as a flow diagram in Figure \ref{fig:flowchart}.
\begin{figure*}
\includegraphics[width=.85\textwidth]{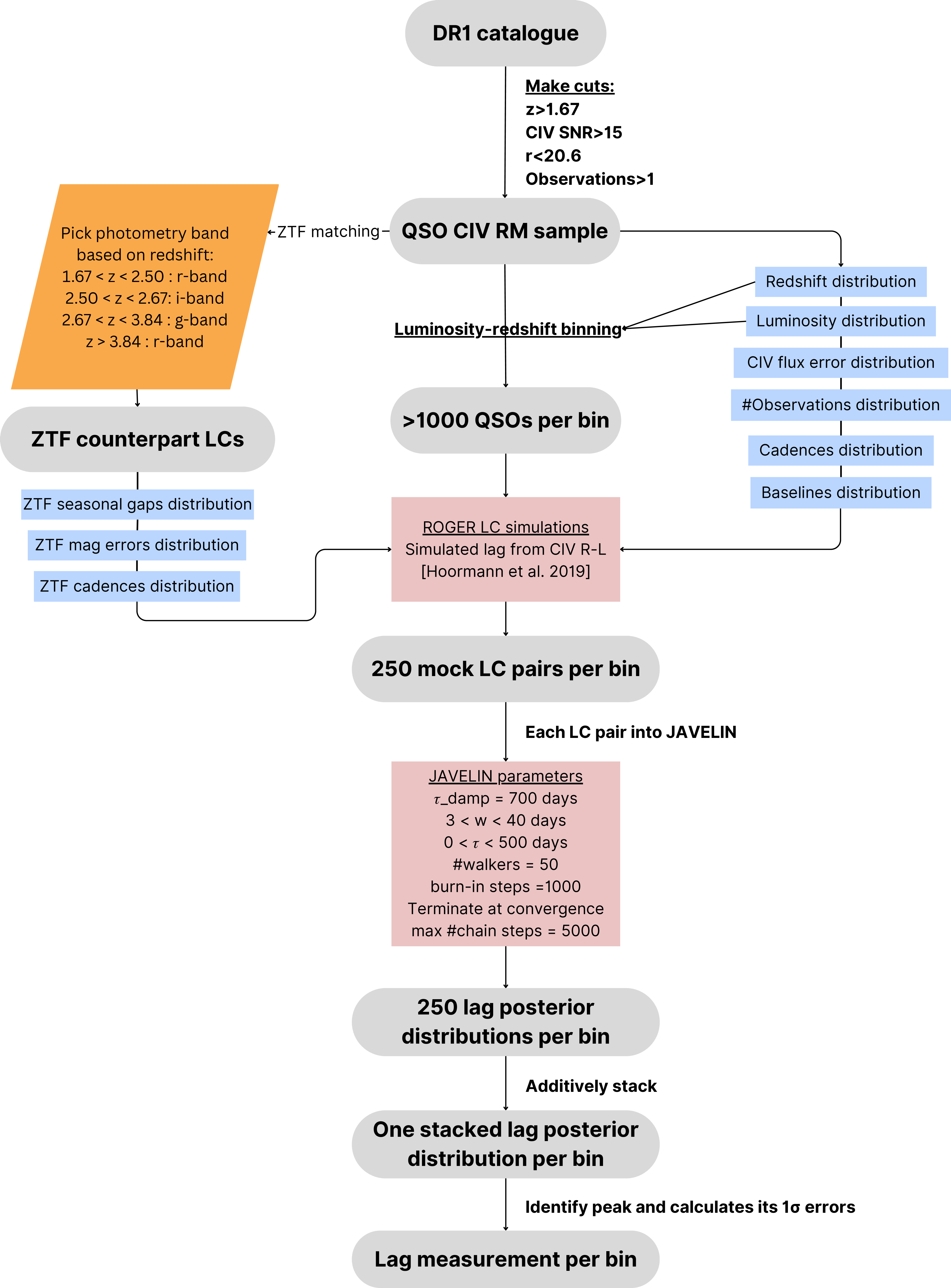} 
\caption{A flow chart summarising the mock stacked RM pipeline. The first half illustrates the various data sets needed to simulate the mock light curves. The second half presents the steps taken to measure a stacked lag from the mocks.}
\label{fig:flowchart}
\end{figure*}

\section{Recovery of the $R$-$L$ relationship} 
\label{sec:results}

We quantify the performance of the stacked RM pipeline introduced in Section \ref{sec:lag_measurement} by examining how well it recovers the input $R$-$L$ relation from \citet{Hoormann2019}, which determined the lag of the mock light curves as described in Section \ref{sec:mocks_method} and outlined in Figure \ref{fig:flowchart}.

The recovered observed lag is transformed to the rest frame and then it is paired with the average luminosity in that bin to give a point on the $R$-$L$ diagram. The measured lags of each luminosity-redshift bin are collectively shown on the $R$-$L$ plane in Figure \ref{fig:rlz}. The majority of lag measurements (96\%) lie within 1$\sigma$ of the input lag, and are therefore successfully recovered. Despite the fact that no single quasar in the mock data sample had sufficient spectroscopic data to detect a lag individually, we have found this stacking procedure allows us to reliably recover average time lags in quasar ensembles for the mock data we have simulated. Thus, it is possible to achieve a $R$-$L$ relation comparable to those obtained by traditional RM methods with data already available in cosmological spectroscopic surveys such as DESI, forgoing the need for observationally expensive RM campaigns to constrain it.
\begin{figure}
\includegraphics{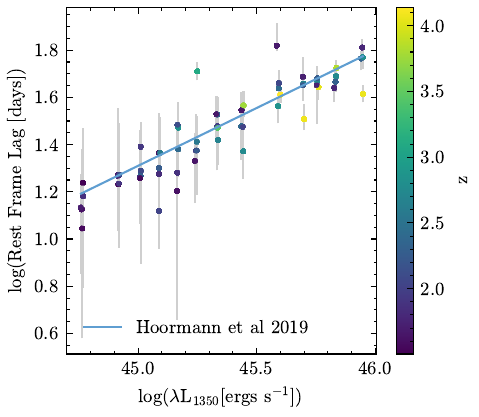}
\caption{The recovered lag of the stacked lag posterior distributions from every luminosity-redshift bin plotted on the radius-luminosity plane. The rest frame lag is equivalent to the light travel time across the BLR. The luminosity of the points is the average luminosity of the objects used in the stack to produce that lag, this is why they roughly appear in columns centred on the middle of the luminosity bins. The grey error bars indicate the 68\% interval of the lag peaks. The points are coloured according to the average redshift of the objects used in the stack. There is no evident bias between redshift and lag recovered. The blue line is the input C~{\sc iv} $R$-$L$ relation from \citet{Hoormann2019} which we are using to test how closely the lags will be recovered.}
\label{fig:rlz}
\end{figure}

The scatter of the recovered lags around the input $R$-$L$ relation is not entirely random. There is a tendency for the lags to be underestimated (see Figure \ref{fig:rlz}). It is worth mentioning that this is the case whether the lag is measured as the MAP or the median of the lag peak, with the median of the lag peak having a greater level of underestimation. This tendency is also found when examining the `outliers': the bins whose lags lie more than 1$\sigma$ away from the true lag. There are seven bins which recovered a lag more than 1$\sigma$ below the true lag. Looking at the lag distributions, these underestimations occurred in noisy distributions with very broad peaks that, although encompassing the correct lag, had their MAP at a lower lag than expected. 

Figure \ref{fig:snr-L} reveals how the SNR of the lag peak degrades with higher luminosities and higher redshifts. This is to be expected as lower luminosity (higher variability) quasars will have stronger light curve features such as maxima, minima, and slopes to match between the continuum and emission line light curves that lead to better lag recovery. For this reason, the outliers tend to be in higher luminosity bins which can also be seen in Figure \ref{fig:rlz}. These simulations demonstrate that when conducting this analysis on real data, the level of scatter or negative offset seen here cannot be discerned from physical effects, such as increased accretion rates, which have been suggested to be responsible for discrepancies in time lags measured in high luminosity quasars \citep{DallaBonta2020, Panda2024}. Since the mock light curves have not been simulated with different accretion rates, we are alerted that this magnitude of scatter is attributed to the lag measurement method as opposed to additional physical parameters that have not been considered.
\begin{figure}
\includegraphics{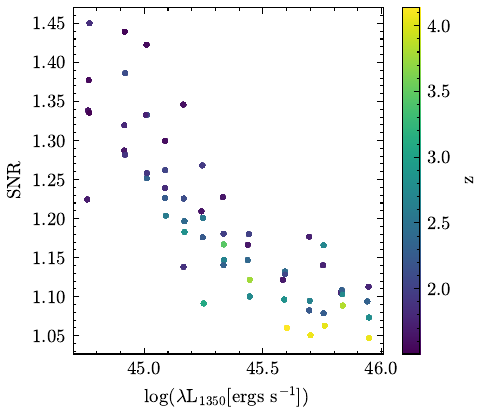}
\caption{The SNR of the lag peak in the stacked lag posterior distributions of every luminosity-redshift bin as a function of luminosity (x-axis) and redshift (colour bar). The SNR has a strong inverse relation with luminosity. There is also an inverse relation with redshift.}
\label{fig:snr-L}
\end{figure}

A best fit of the recovered lags tests whether the input $R$-$L$ relation we simulated the lags with is successfully recovered. This is done by averaging the lags in each luminosity bin and using the range of recovered lags for that bin as the statistical error (as seen in Figure \ref{fig:rl}). The recovered $R$-$L$ relation is obtained by a least-squares line of best fit in the log-log plane that is weighted by the average SNR per bin. The final recovered C~{\sc iv} $R$-$L$ relation is
\begin{equation}
\log R \text{[days]} =  (0.78\pm 0.035) + (0.51 \pm 0.025)\log\frac{\lambda L_\lambda(1350\text{\AA{}})}{10^{44}[\mathrm{erg\,s^{-1}}]},
\label{eqn:recoveredRL}
\end{equation}
which is in agreement with the input relation given in Equation \ref{eqn:HoormannRL} to 1$\sigma$. 

In Appendix \ref{app:lagsgaps}, we present additional tests to check the robustness of the pipeline by modifying the light curves to have no lag and/or no seasonal gaps. These tests did not indicate that the bias towards lower lags is caused by seasonal gaps but rather this underestimation is due to some other systematic factor. 

\begin{figure}
\includegraphics{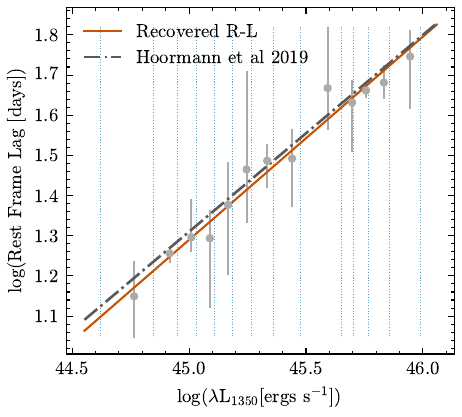}
\caption{The final radius-luminosity relation recovered from the mock DESI quasar light curve suite is shown as a solid red line. It is to be compared to the grey dot-dashed line taken from \citet{Hoormann2019} used as an input for the mock pipeline. The grey points are the average of the recovered lags in each luminosity bin (indicated by dotted blue vertical lines). Their error bars reflect the range of recovered lags in that luminosity bin across all redshifts.}
\label{fig:rl}
\end{figure}

In the development of the stacked RM pipeline, we identified several factors that, if different, could potentially effect the lag recovery. In the subsections below we systematically investigate the dependence of the final result on each of the following factors:
\begin{itemize}
    \item spectroscopic C~{\sc iv} flux errors,
    \item photometric light curve baseline,
    \item photometric seasonal gaps,
    \item number of quasars stacked per bin,
    \item number of spectroscopic observations per quasar, 
    \item and bin width.
\end{itemize}
In the conduction of these tests, we have used the same suite of underlying light curves as the base run presented in this section and the findings are compared to the result above. We conduct these tests by changing one factor at a time rather than compounding factors, making the source of any change clear.

\subsection{Dependency on the C~{\sc iv} Flux Errors}
\label{sec:flux_err}
First, we test how multiplying the C~{\sc iv} integrated flux error distribution by factors of 0.5, 1, and 2 will effect the recovered $R$-$L$ relation. The inflation of the error distribution affects both how far the sampled epochs deviate from the underlying light curve during mock light curve generation and the magnitude of the errors assigned to the data points in the resulting light curve.
From this test, we aim to verify whether the quality of DESI spectra is sufficient for stacked RM. The error of the integrated flux will ultimately depend on the exposure time of that spectral epoch. This is not something that can be easily controlled so we test both scenarios of worsening and improving the data quality around the current DESI standard to see what weight the error factor has on the final result.

We find that our pipeline is not too sensitive to flux error inflation as all three flux error factors reproduce the input $R$-$L$ relation well. The differences in the $R$-$L$ line fits are within the bounds of random variation between runs. The impact of the flux error becomes apparent when analysing the SNR of the lag measurements, as shown in Figure \ref{fig:fluxerrs}, the smaller the flux error the closer the measured lag is to the input lag and the higher the SNR. Although this trend is present, the majority of measurements are below the 1$\sigma$ line shown in grey and so we find that for the error factors explored here, no significant issues or differences arise from the pipelines results. \citet{Yu2020} found that {\sc javelin}'s use of Gaussian processes makes it more sensitive to light curve errors than other lag measurement methods \citep{Aigrain2023}. Therefore, the calculation of accurate flux uncertainties and possible routes of minimising it, either by way of sample selection or by using a method that calculates it more precisely, is important. Nevertheless, this stacked RM pipeline has an advantageous feature that the final $R$-$L$ relation is well-recovered within the range of typical flux errors found for the DESI data.
\begin{figure}
\includegraphics{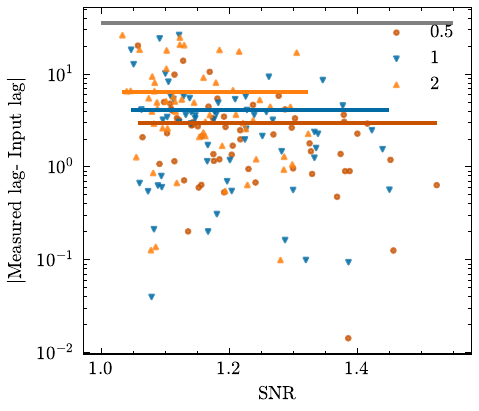}
\caption{The absolute difference between the measured lag and the input lag of a bin against the SNR of the lag peak in the bins stacked posterior distribution. The results are shown for three runs in which the spectroscopic flux errors are multiplied by the factors: 0.5 (red circle), 1 (blue inverted triangle), and 2 (orange upright triangle). The height of the horizontal lines mark the average deviation from the input lag and the span indicates the range of SNR's in that run. Reducing the line flux errors results in more accurate lag recoveries and enhances the SNR of the lag peaks. The gray line indicates the median 1$\sigma$ lag error of the \citet{Hoormann2019} relation across all bins to illustrate all three runs are well within 1$\sigma$.}
\label{fig:fluxerrs}
\end{figure}

\subsection{Dependency on the Photometric Light Curve Completeness}
\label{sec:phot_compl}

\begin{figure}
\includegraphics{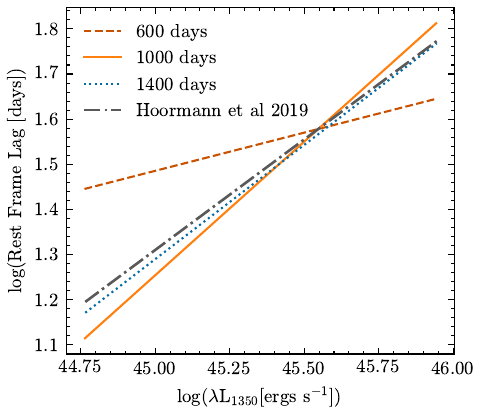}
\caption{$R$-$L$ lines recovered for three runs in which the photometric light curves spanned a baseline of 600 (dashed red), 1000 (solid orange) and 1400 (dotted blue) days. 1400 days is what is used in the base run. This plot clearly demonstrates the drastic sensitivity that this method has to the photometric baseline. A shorter baseline makes recoverability significantly worse.}
\label{fig:photbas}
\end{figure}

\begin{figure}
\includegraphics{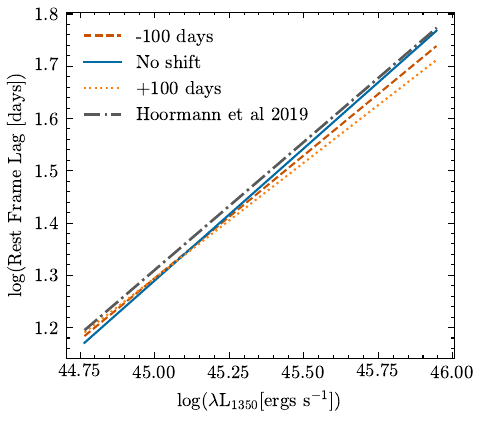}
\caption{ $R$-$L$ lines recovered for three runs in which the distribution of seasonal gap lengths is shifted by -100 days (dashed red), +100 days (dotted orange) and the original with no shift (solid blue). The method seems fairly resistant to the narrowing and widening the seasonal gaps.}
\label{fig:seasongaps}
\end{figure}

\begin{figure*}
\includegraphics[width=\textwidth]{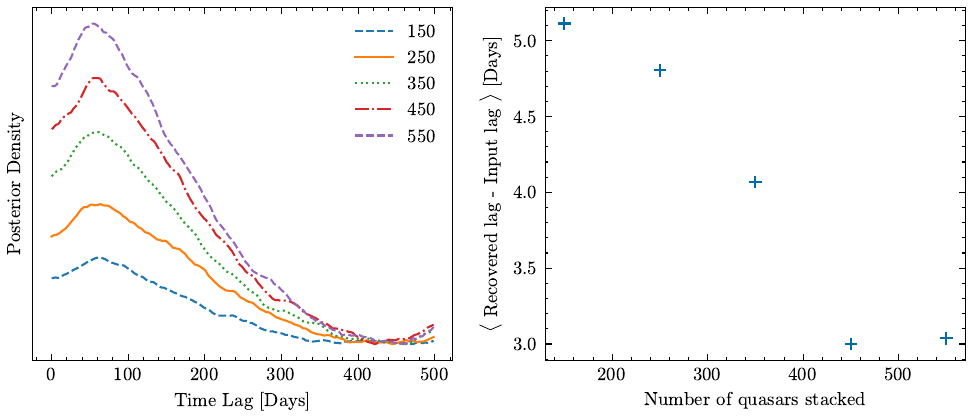}
\caption{Left: Evolution of the averaged stacked lag posterior distribution as the number of objects stacked is increased from 150 to 550. The distributions have been translated so their minima align. As the number of objects stacked in the bin is increased the peak becomes more prominent, i.e. the SNR increases. Right: The evolution of the average deviation of the recovered lag from the input lag with the number of quasars stacked. There is an inverse relation below $\sim$450, beyond this the improvement in the lag measurement plateaus.}
\label{fig:numstack}
\end{figure*} 

In the creation of our mock photometric light curves we picked a set baseline of 1400 days which reflects the average data available from ZTF. We probe the underlying mock light-curves with different window-functions to investigate the impact it has on lag recovery. We investigate how using shorter photometric baselines prior to the spectroscopic observations affects our results by additionally testing two alternative baseline lengths of 600 and 1000 days. Figure \ref{fig:photbas} shows the $R$-$L$ relations each of these runs produced. There is some deterioration in recovery from 1400 days to 1000 days of photometric data available. The scatter of the recovered lags around the input $R$-$L$ relation for the main run with a 1400 day long photometric baseline is 0.033 dex and this doubles to 0.074 dex for a 1000 day baseline. Once we restrict the photometric data to 600 days the ability to recover accurate lags severely declines, with much lower SNR lag peaks in the stacked posterior distributions. The scatter of the recovered lags spikes to 0.29 dex, almost a factor of 9 larger than the main run. This can be attributed to the fact that the first step of {\sc javelin} is to characterise the variability amplitude of the continuum light curve. When we restrict the photometric light curve data to range a shorter span of time we are limiting the accuracy of the variability analysis of the quasar and this carries forward to poor lag measurement. This is to the extent that out of the 250 quasars only 45 lag measurements are (poorly) recovered. The rest failed at the first step of {\sc javelin} which was unable to characterise the truncated continuum light curve and the walkers diverged.

We also investigate changes in the completeness of the photometric light curve from the angle of the length of seasonal gaps. As mentioned in Section \ref{sec:lcsim}, we simulate seasonal gaps using a distribution of their length in days from ZTF counterpart LCs of DESI quasars. We shift this distribution by -100 days, which creates light curves with short seasonal gaps and high completeness, and +100 days, which creates light curves with long seasonal gaps and low completeness. Figure \ref{fig:seasongaps} shows how the recovered $R$-$L$ relation changes with these three runs. We do not find a trend between the length of seasonal gaps and the recoverability of the $R$-$L$ relation. Even the greatest deviation from the input $R$-$L$, seen when increasing the length of seasonal gaps, is but a slight deterioration beyond the random variation expected between different runs. However, we did see a trend in SNR which increased with more complete light curves.

\subsection{Dependency on the Number of Quasars Stacked} \label{sec:numStack}

The number of quasars per stack in the main run is set to 250. We investigate how the number of objects stacked effects the SNR of the lag peak in the stacked distribution. We tested this by running the pipeline five times changing the number of objects stacked to 150, 250, 350, 450, and 550 quasars per bin. Figure \ref{fig:numstack} shows the average stacked lag posterior distribution for each run. The clear trend is that the more objects that are stacked the more defined the time lag peak is, i.e. the higher the SNR. When comparing the recovered lags to the simulated lags we also see that accuracy in recoverability steadily increases with the number of objects stacked. Figure \ref{fig:numstack} shows that the average deviation of the recovered lag from the input lag declines up to a stack of 350 quasars, but beyond this the improvement plateaus. In summary, increasing the number of stacked objects aids in the recoverability of the lag since the lag peak in the distribution is further enhanced. All of the $R$-$L$ lines recovered for each of these runs are within 1$\sigma$ of the input relation, so ultimately the recovered $R$-$L$ relation does not differ greatly whether it is with 150 or 550 objects per bin. However, in order to reduce the scatter and therefore the error in the final $R$-$L$ relation fit, a stack with at least 400 quasars is best.

\subsection{Dependency on the Number of Spectroscopic Observations} \label{sec:extra}
In the inaugural stacked RM paper, \citet{Fine2012} predicts that this procedure can be inclusive of quasars which only have two spectroscopic observations. In the main run, $\sim 1/3$ of the 250 quasars per bin had two observations and the remaining had more than two. The inclusion of such a large fraction with as few as two observations solidifies the value of these quasars in stacked RM. Given DESI's vast number of unique quasar observations, it may be possible to place tighter cuts on the stacked RM quasar sample to only include quasars with more than two spectral epochs. We investigate how changing the minimum number of spectroscopic observations permitted may further improve lag measurements. We tested limiting the quasar light curves to having $\geq4$ and $\geq 8$ spectroscopic data points. In terms of lag measurements, Figure \ref{fig:numspec} shows how increasing the number of repeat observations will draw the recovered lag closer to the input lag. We also see that having $\geq8$ spectroscopic points can even triple the SNR. 
Given that the $R$-$L$ recovery is once again within 1$\sigma$ in all three runs, stacked RM is not only effective with the use of poorly sampled spectroscopic light curves but, considering their exclusion may limit the number of quasars stacked for a DESI-like survey, these objects should be included.
\begin{figure}
\includegraphics{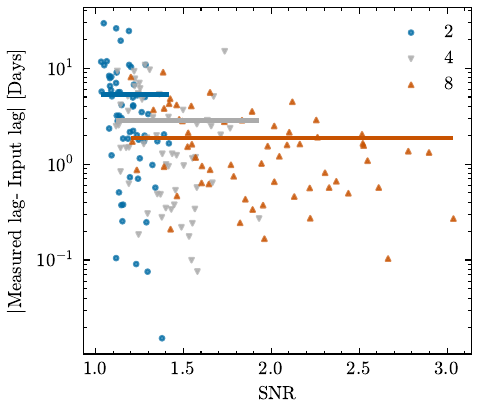}
\caption{The difference between the measured lag and the input lag of a bin against the SNR of the lag peak in the bins stacked posterior distribution. The results are shown for three runs in which the minimum number of spectroscopic points in the broad emission line  light curve is increased from the base run with $\geq2$ (blue circle) observations up to $\geq4$ (grey inverted triangle), and $\geq8$ (orange upright triangle) observations. The height of the horizontal lines mark the average deviation from the input lag and the span indicates the range of SNRs in that run. It is clear that providing a larger number of spectroscopic observations results in more accurate lag recoveries and bolsters the SNR of the lag peaks.}
\label{fig:numspec}
\end{figure}

It is pertinent to explore the baselines that these spectroscopic observations span. Figure \ref{fig:bascad} investigates how varying the maximum spectroscopic baseline in the mock light curves alongside the minimum spectroscopic observations influences the lag recovered. This is tested via 16 runs on a singular luminosity-redshift bin varying only these two factors. The result clarifies that accurate lag recovery is more strongly a function of spectroscopic baseline as opposed to the number of spectroscopic epochs. This is a key result emphasising that stacked RM programs with large spectroscopic surveys should be designed to prioritise long baselines. It is observationally more efficient to obtain a small number of spectroscopic observations spread over a few years, rather than collecting multiple epochs within a single year that still fail to yield a comparably accurate time lag.

\begin{figure}
\includegraphics{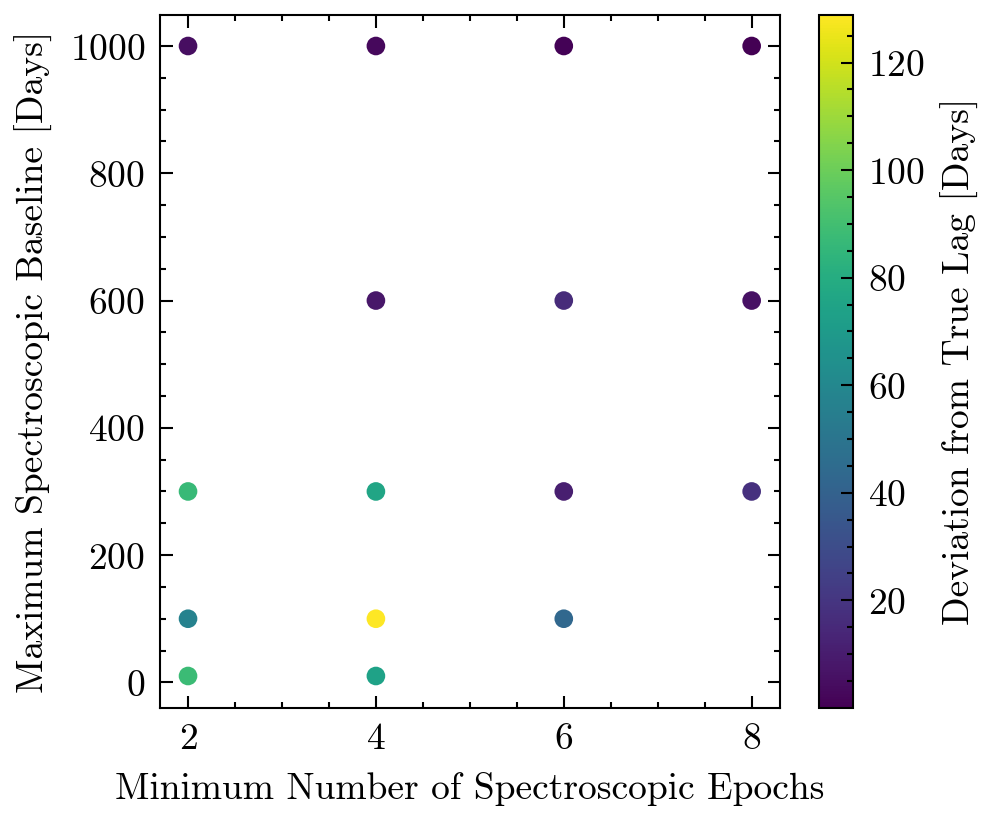}
\caption{Absolute deviation of the recovered lag from the input lag as a function of the minimum number of spectroscopic observations and the maximum baseline set in generating the mock light curves. The results of the 16 runs shown are for the same luminosity-redshift bin with an average luminosity of  $2.1 \times 10 ^{46} \mathrm{erg\,s^{-1}}$ and an average redshift of 3.2. Quasars with spectroscopic light curves whose  epochs occur within a short period of time are unable to recover the true lag, no matter the number of epochs. It is required for the stacked sample to contain spectroscopic light curves with a baseline that is at least a few hundred days long, i.e. on the scale of the observed lag, for a successful lag detection. A higher number of spectroscopic epochs increases accuracy in the lag recovered, however, the extent of the spectroscopic baseline is a stronger driving factor of lag recovery.}
\label{fig:bascad}
\end{figure}

\subsection{Dependency on the Bin Size}
\label{sec:binSize}

\begin{figure}
\includegraphics{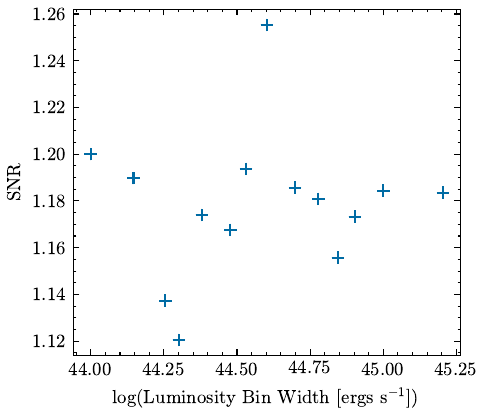}
\caption{The SNR of the lag peak in the stacked posterior distributions as we increase the width of the luminosity bin. A strong correlation is not present.}
\label{fig:binsize}
\end{figure}

The creation of 61 luminosity-redshift bins in Section \ref{sec:binning} is designed to provide a reliable stacked lag result. The bin widths are arbitrarily based on equal-population binning. Here, we experiment with changing the luminosity bin width. We fixed the redshift bin to be $2.0\leq z \leq 2.2$ and created luminosity bins centred at $2 \times 10 ^{45} \mathrm{erg\,s^{-1}}$ with bin widths ranging from $1 \times 10 ^{44}$ to $1.6 \times 10 ^{45} \mathrm{erg\,s^{-1}}$. When applying wider binning to data, a larger and more diverse sample of quasars contribute to the stack. Since the number of quasars and the diversity of quasars counteract each other in the ability to recover the lag, in this mock test we have set the number of quasars fixed to 250 to isolate the effect of having a broader range of lags contributing to the stack. Figure \ref{fig:binsize} aims to explore whether there is a relation between the bin width and the SNR of the lag peak. Surprisingly, we do not find a strong correlation between the SNR and the width of the luminosity bin. There is a shallow slope towards the expected trend that larger bins sizes reduce the SNR of the lag peak - this is due to the inclusion of a more diverse range of quasars and therefore lags, widening the lag peak. To note, the recovered lags are all within 1$\sigma$ of the input lag. 
For surveys with a smaller sample of quasars, this finding alleviates the requirement for narrow luminosity-redshift bin widths. If luminosity-redshift precision is not needed for a RM study, this pipeline is suitable for RM in a low resolution luminosity-redshift space. This leads to a greater emphasis on having bins with a larger population of quasars.

\section{Robustness of Stacked RM} \label{sec:discussion}

We have demonstrated via mock data that stacked RM is a viable approach to determining the $R$-$L$ relation for quasars observed by DESI. Whereas traditionally the continuum and emission line light curves would need to overlap for a period that is at least a few times the time lag to obtain a lag measurement, we have shown this is unnecessary for calculating lags for ensembles of similar quasars, confirming the proposal of \cite{Fine2012}. Well-sampled spectroscopic light curves of individual quasars are observationally expensive. We have demonstrated here that the same result, the same $R$-$L$ relation constructed from a small group of quasars, can be obtained at a small fraction of the observational cost as a by-product of the multitude of quasar observations conducted by a large spectroscopic survey such as DESI. The benefit of conducting a RM study using a composite sample of thousands of quasars is that a statistical approach provides a more robust and unbiased means of inferring the $R$-$L$ relation, enhancing its reliability for single-epoch black hole mass estimates and applications in quasar cosmology. The merging of quasar observations in the stacking procedure has the useful effect of averaging out some of the unknowns related to the geometry of the system such as the range of inclination or viewing angles within the sample. We have specifically demonstrated this pipeline's success for high redshift quasars, a regime which currently has a dearth of reliable observational results. This proof of concept is not limited to high redshifts or RM with the C~{\sc iv} line. It can be used for other broad emission line species, such as H$\alpha$, H$\beta$, and Mg {\sc ii} as well as lines which have not received as much attention such as He {\sc i}, He {\sc ii}, and H$\gamma$. An investigation into the sensitivities of this method has laid out a clear picture of the boundaries of the data suitable for this pipeline.

\subsection{Limitations of Pipeline} \label{sec:limitations}

There are several possible limitations of the pipeline which we discuss here.
RM is limited by an artificial lag cut-off induced by the combination of a finite observational baseline and lag range prior which, in effect, flattens the $R$-$L$ relation at high luminosities. In this study the highest lag simulated is 383 days in the observed frame so the lag prior upper boundary of 500 days is not impinging on the lag recovery. Our photometric baseline extended to 1400 days so the lag measurement is not restricted on this front either. We made the decision to not include the highest luminosity bin quasars (see Figure \ref{fig:lumzdistbins}) due to insufficient numbers in that range to allow for proper binning. This area of the luminosity-redshift space may contain quasars with lags that go beyond 1000 days in the observed frame. If these are to be included in future studies, it is imperative that the lag prior is expanded, perhaps to 1500 days, and that the photometric baseline is extended comparably.

The $R$-$L$ relation has an intrinsic scatter in addition to the dispersion due to uncertainty in luminosity and time lag measurements \citep{KilerciEser2015}. Stacked RM may be vulnerable to the dispersion in the $R$-$L$ relation due to the inclusion of a more diverse collection of quasars which is necessary due to the size of the sample needed. The outcome will be that the averaged result will be biased towards the subset of the quasar population with higher variability patterns at a given luminosity since they have stronger lag signals. Past studies have shown limiting the sample to a specific subset of the quasar family decreases the $R$-$L$ relation dispersion, such as placing an upper limit on the accretion rate \citep{Du2019}.

The transfer function that transmutes the continuum light curve to the emission line light curve (see Equation \ref{eqn:transfer}) is assumed by {\sc javelin} to be of a certain parametric form, the top-hat transfer function given in \citet{Zu2011}. There have been other parametric forms for the transfer function investigated such as Gaussian or exponential \citep[e.g.,][]{Li2016, Yu2020}. When conducting this pipeline on real light curves it is worth bearing in mind that the top-hat transfer function is very simplistic given that the transfer function may be asymmetric and multimodal. 
By using {\sc javelin} in our mock pipeline, we are assuming the same quasar variability and light curve model as those assumed in the light curve simulations. The results from this mock study here are free from the issues caused by theoretical simplifications in a data study. It is left to future work to investigate the effect of simulating the light curves with transfer functions of different parametric forms on the stacked RM results. 

The simulations have also assumed that the broad emission line light curve will always be echoing the continuum variations throughout the observations baseline. In reality, this may not be the case as \citet{Dehghanian2019} discovered that during the intensive monitoring campaign of NGC 5548 by the AGN Space Telescope and Optical Reverberation Mapping project (AGN STORM, \citealt{Derosa2015, Horne2020}) there was a period during which the broad emission lines became decoupled from the continuum and did not exhibit the reverberations found at other times. This was coined the `BLR holiday'. When the BLR is on holiday, the broad emission line light curve does not show any correlation to the variations in the continuum light curve and hence does not display any lag. The frequency of BLR holidays and the fraction of quasars in the stack that have DESI epochs amid such a holiday may have a significant impact on the lag measurement as these objects in the stack will only add noise without a signal. This requires further investigation.

One of the greatest difficulties in this project was the employment of {\sc javelin}. The {\sc javelin} code was originally created for a specific purpose and to conduct RM for a particular set of data. It is a great resource for measuring lags according to a Bayesian framework, yet, it is inherently designed for well-sampled light curves of a certain nature. For example, the ratio between the time lag and the length of the light curve is an important factor for {\sc javelin}. The light curve should be several times longer than the lag, whilst the continuum light curves meet this criteria, the C~{\sc iv} light curves are often shorter or comparable to the lag. In the process of developing the pipeline as described in Section \ref{sec:lag_measurement}, we ran into many numerical issues that resulted in the MCMC chains not being able to converge with the mock data provided. The pipeline described herein represents the approach we ultimately identified as effective for addressing the challenges posed by our sparse dataset. We wish to highlight the sensitivity of the {\sc javelin} parameters we have defined, as they may significantly influence the results. It should also be noted that {\sc javelin} may not be the most practical or universally applicable tool for all RM studies. In light of this, we encourage the exploration of stacked RM with other lag measurement codes that are currently available or are being developed. For example, the {\sc litmus} code was specifically developed to overcome the numerical issues that frequently occur in {\sc javelin} \citep{Mcdougall2025}.

There are numerous studies questioning the validity of the DRW model for variability assumed in this project. In this project, we have confirmed that, with the assumption of DRW-abiding light curves, stacked RM can be successful with {\sc javelin} or perhaps another Bayesian code which assumes DRW. Studies show that the DRW assumption does not hold universally, for example for high and low frequency variations \citep{deVries2005, Kozlowski2016, Read2020, Yu2022, Stone2022}. DRW does not take into account flares, reprocessing, and feedback. C~{\sc iv} is often associated with or affected by outflows which may mean its variability pattern can significantly deviate from DRW. Therefore, interpolating the observed light curve upon this model is a potential source of error and can produce false lag peaks. The stacking procedure wards off these false lags as the real peak is bolstered, but assuming a more correct model would help strengthen it even more. The alternative options are to either adopt a more complicated model and still reap the benefits of a Bayesian analysis or to adopt lag measurement techniques that are independent of a variability model. CCFs are an example of the latter approach, and they may serve as a useful consistency check, despite their known limitations, as we have done in this study in Section \ref{sec:CCF}. To assess potential deviations from a DRW model, we introduced noise into the observations time axis to perturb them away from ideal DRW behavior. This modification did not result in any significant changes to the outcomes. In future work, it can be tested further by simulating the light curves with a completely different model or by using empirical realisations of the DESI-ZTF light curves to see if the lags can still be recovered by {\sc javelin}. So far, this mock pipeline suffers from the concern of possibly unrealistic light curve variation simulation. 

When we are conducting the time lag measurement with {\sc javelin}, we define uniform priors on the parameters. In the stacking procedure we have made an implicit assumption that the priors of the time lag parameter are independent. \citet{Brewer2014} counsels that it may be incorrect to infer a parameter from its distribution across a sample where the individuals are analysed separately yet have dependent priors. We have purposefully designed our method such that the quasars that are stacked have time lags clustered around the average time lag for that luminosity-redshift bin, hence, the priors are not independent. Accordingly, \citet{Brewer2014} presents a method named hierarchical reverberation mapping whereby the use of hyperparameters are used to overcome this issue. Whether the issue is genuinely a cause for concern is not clear since we have found that a simple additive stacking competently recovered lags. Hierarchical RM may be more computationally expensive given the additional parameters but it does have the benefit of producing more precise lag measurements and making the interpretation of the peaks more straightforward.

We do not recommend any quality cuts for the lags produced by this pipeline since we did not find any suitable SNR cut-off for the lag peak which distinguishes the accurately recovered lags. An example can be seen in Figure \ref{fig:numspec} where there are many low SNR lags which accurately match the input lag and, on the other hand, high SNR lags are not necessarily more accurate compared to low SNR lags. The distinguishing factor between low SNR and high SNR lags is high and low luminosity quasars, as shown in Figure \ref{fig:snr-L}. This effect is also present at a given redshift. Although at a given redshift high luminosity quasars have higher SNR epochs in their spectra, the variability is still lesser than those of low luminosity quasars, and therefore they have a weaker lag signal. In a way, it is advantageous that the low SNR lags are still able to recover the lag accurately as enforcing a SNR cut would imply a luminosity cut on the sample and the aim of this study is to expand into the high luminosity regime. 

One of the findings of this study is that the developed pipeline tends to underestimate lags. As confirmed through CCF analysis in Section \ref{sec:CCF}, this underestimation appears consistent across lag recovery methodology, indicating that its source likely originates earlier in the pipeline. We have not found any correlation between the magnitude of this underestimation and parameters such as redshift, luminosity bin width, or redshift bin width.
We observe that the systematic underestimation of lags arises from the shape of the lag distributions, which are skewed toward shorter lags rather than being uniformly flat with a superimposed peak. This phenomenon is evidenced by higher noise levels on the short-lag side of the peak in the stacked distributions (this can be seen in Figure \ref{fig:stackedpd}). Figure \ref{fig:avpd} is of the averaged posterior distribution across all bins and it clearly illustrates this overdensity. The average lag across all bins is 108 days. However, when we average the stacked posterior distributions of the bins, the peak does not align with the average lag, its MAP is positioned at 55 days. This considerable discrepancy persists even in the presence or absence of seasonal gaps, as shown in Appendix \ref{app:lagsgaps}. The stacking process amplifies not only the main peak but also the low-lag overdensity, introducing a negative slope in the stacked distribution. Consequently, shorter lags are overrepresented, biasing the overall result. Low luminosity quasars, which naturally have shorter lags, are especially susceptible to this effect.

\begin{figure}
\includegraphics{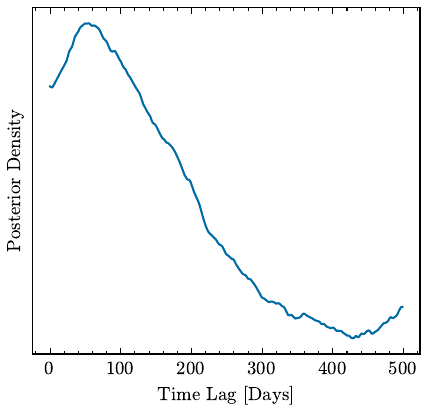}
\caption{Average of all the posterior distributions across all time-lag bins. The peak is shifted to shorter lags (maximum at 55 days) in comparison to the expected average of the peaks of the stacked posterior distributions (maximum at 108 days). This is a manifestation of the systematic overdensity at the lower lag end present in most of the stacked lag posterior distribution. The underlying distribution beneath the peak is not uniform as expected but there is a negative slope.}
\label{fig:avpd}
\end{figure}

The SDSS stacked RM study presented in \citet{Li2018} found something similar with stacked zDCFs producing average lags of the sample significantly shorter than the H$\beta$ $R$-$L$ relation derived from individual lag measurements by \citet{Bentz2013}. They proposed that this could be due to two reasons. Firstly, since their stacked lag contained a wide range of luminosities and redshifts they believed that the average lag was dominated by the lower redshift and lower luminosity quasars which would have stronger lag signals given their higher variability. They believed the solution to this would be to weight the stacking based on the errors of individual zDCFs. The second reason they suggested for the lower than expected lags is that the sample contained quasars with higher Eddington ratios than the local sample used to construct the \citet{Bentz2013} $R$-$L$ relation and therefore would have a lower average lag. Our study here eliminates both of these reasons for potential causes of the underestimation. Our stacks of quasars have a sufficiently small range of lags per bin to discount the diversity of quasars as being the reason for underestimation. For comparison, the average standard deviation of the rest lags in a bin is around 1 day whereas the standard deviation of the scatter around the expected $R$-$L$ relation is 6.5 days. 
The second reason is disproved by our light curve simulations in which the true lag is known so the underestimation of the recovered lag is literal and not due to sample differences. The OzDES-RM stacked RM analysis carried out in \citet{Malik2024} did not encounter this underestimation. They even investigated the discrepancy between the shorter SDSS-RM lags and their stacked lags consistent with \citet{Bentz2013} via simulations of the differences in baseline and cadence and did not find that the difference in observational programs led to shorter lag measurements. In this study, we have further verified that the underestimation in lag is not linked to the presence of seasonal gaps, and since the baseline was a controlled factor, it is not due to this either. The stacked RM analysis by Eltvedt et al. (in prep) also displayed consistently lower stacked lags than those predicted by literature $R$-$L$ relations derived from individual lags. This was dismissed since their large uncertainties meant that the lags are still in agreement.

It is still unclear exactly what is causing the overdensity at the lower end of the lag posterior. Some of the previous studies performed stacked RM on light curves with far better spectroscopic sampling so it is unlikely that the sparse cadence is the culprit. We verified this using the runs with increased spectroscopic observations described in Section \ref{sec:extra}, and found that improving the cadence did not affect the proportion of underestimated lags.
Further investigation into this phenomenon is critical for characterising scatter in the $R$-$L$ relation so that we can decompose the scatter into numerical and physical components.

\subsection{Limitations in Observations} 
\label{sec:limit_obs}
In this mock study, we did not need to circumvent obstacles related to observations. Here, we briefly discuss the array of observational limitations one must tackle when applying this pipeline on real data.

It is likely that photometric bands may still be contaminated with broad line flux even after the process we described in Section \ref{sec:mocks_method}. We have only taken into account the most dominant broad lines in our target redshift range, C~{\sc iv} and Mg {\sc ii}, but others may be present such as C {\sc iii}] $\lambda$1909\AA{}. Some high BH mass quasars or quasars with outflows may have very broad lines that extend beyond the width we used to calculate the redshift ranges for the photometric bands and so would leak into the neighbouring bands. This part of the quasars spectrum also covers the Fe {\sc ii} pseudo-continuum \citep{Kovacevic2010} and diffused scattered BLR emission \citep{Korista2019, Pandey2023, Jiang2024}. All non-continuum contaminations have the power to alter the BLR lag measurement. The photometric light curves may also need to be "detrended", this is the process of removing long time-scale secular variations that are unrelated to reverberation \citep{Welsh1999, Li2013, Peterson2018, Edelson2024}. 

RM also makes the assumption that the BLR is stable, i.e. the time lag is constant. The time lag may not be observed to be constant due to obscuration effects. There is also the possibility of extra sources of broad line variability and variability due to microlensing \citep{Hawkins1997}. Additional non-reverberating components that effect the shapes of the light curves can lead to underestimation of lags. One could strive to purify the light curve of these extra components. Quasars are complex systems and simplifying assumptions are made to gain an estimation of the size of the BLR. As better data becomes available, RM will be able to probe finer details of the BLR environment. For example, past studies have investigated BLR geometry via velocity-resolved RM \citep{Bentz2008, Denney2009, Villafana2012, Bentz2023}. However, stacked RM provides relief from some of these concerns as it does not claim to achieve individual lags. It offers an average lag that blurs out the secondary lag influences caused by the complexities and diversity of individual systems.

The main reason behind high luminosity-high redshift bins having worse lag measurements in terms of SNR is due to low variability and time dilation. To capture these quasar's variability patterns well, it is necessary for the observational cadence to be spread out further and extended beyond the cadence of DESI DR2 data used in this mock study. The DESI data used in this study is sufficient to recover lags for these quasars but, to increase the SNR, follow-up observations are needed to supply more spectral epochs in order to produce the same quality measurements as the low luminosity-low redshift quasars (see Figure \ref{fig:numspec}). If these cannot be acquired then the survey design's constraints can be counteracted by stacking more quasars in the higher luminosity-redshift bins.

If one is to take the approach of having very narrow luminosity-redshift bins, the observed lags being stacked will be very similar. Seasonal gaps cannot be avoided. Extra care is needed to check whether the lag for that bin would not mean the spectroscopic data always falls into the same photometric seasonal gap making lag measurement unsuccessful. 

We placed a broad line detection cut on the quasar sample of C~{\sc iv} flux SNR > 15. This is implicitly biasing the sample to brighter quasars and causes the recovered $R$-$L$ relation to be dominated by lower variability sources.

RM is only possible when there is detectable intrinsic variability in the light curves. We tested the variability of the quasar sample using the photometric light curves by calculating the variability factor defined 
\citep{RodriguezPascual1997, Bao2022, Goncalves2025} as:
\begin{equation}
    f_{\text{var}} = \frac{(\sigma^2-\Delta^2)^{1/2}}{\langle f\rangle }, 
\end{equation}
where
\begin{equation}
    \Delta^2 = \frac{1}{N} \sum^{N}_{i=1}\Delta_{i}^2,
\end{equation}
$\langle f\rangle$ is the unweighted mean flux, $\sigma$ is the rms flux, $N$ is the number of individual observations, and $\Delta^2$ is the mean square values of the uncertainties. All the quasar light curves in our data sample had detectable variability, i.e., $f_{\text{var}}>0$. This reassures us that RM is viable for even the highest luminosity and highest redshift bins, which are the least variable. We repeated this analysis on our mock light curves which gave a variability distribution very similar to the variability distribution of the photometric light curves; yet another confirmation of the soundness of the simulated light curves.

\subsection{Future Uses}

This pipeline can be fine-tuned in future studies in a myriad of ways. One way is to use {\sc javelin}'s functionality for simultaneously fitting lags with two broad lines which may help eliminate false correlations between light curves and prevent aliases in the lag distributions. Similarly, two photometric bands instead of one, if available, may be used by {\sc javelin} for characterising the quasars continuum variability with higher certainty. 

Another opportunity of research is to exploit DESI's large and diverse quasar sample and further split it by various observational and intrinsic properties in order to elucidate how the $R$-$L$ relation varies with quasar sub-populations. This split could be in accretion rate, looking at sub-Eddington, Eddington, and super-Eddington accretion  \citep{Du2019, Donnan2021} or simple empirical cuts to define bright and faint samples (Eltvedt et al. in prep). 

Stacked RM has a promising future ahead. There are several opportunities within upcoming surveys to supplement the spectroscopic and photometric data explored in this paper. This includes the additional data obtained for the remaining durations of both DESI and ZTF surveys. Our results in Section \ref{sec:extra} demonstrates the benefits of increasing the number of spectroscopic observations and baselines. There may be opportunity to add follow-up observations of DESI quasars through a secondary target program. At the finalisation of the DESI survey, it will be followed by the DESI-II program which aims to bring the accomplishments of DESI to the higher redshift regime, since DESI-II will include additional observations of at least the $z>2$ quasars \citep{Schlegel2022}. This will be of key importance to advancing the path laid out in this work with the availability of larger samples of quasars with $\geq2$ observations to constrain the $R$-$L$ relation. It is also possible to append past spectroscopic data from surveys such as SDSS. Other data opportunities that may improve the results obtained by stacked RM could be a dedicated program using WEAVE \citep{Jin2022}, or the up-and-coming 4MOST's Time Domain Extragalactic Survey \citep[TiDES][]{Frohmaier2025}, or WST \citep{Bacon2024}, or possibly MOONS for infrared RM \citep{Cirasuolo2020}. These will all serve to increase the sample size available for more complex stacked RM studies and increase the number of spectroscopic observations per quasar. On the photometric side, great advances will be made with the advent of LSST which will provide 10 years worth of high cadence photometry.

Among the benefits of the investigation described in this paper is that we state the baseline data required for a successful stacked RM program with DESI to be the following. There should be at least 400 quasars per bin. A lag signal is not detectable with only tens of quasars but it should be with hundreds. Any quasar with $\geq$2 observations should be included. Quasars with only two observations need not be discarded in favour of quasars with more observations, retention of a larger sample is more important. Nonetheless, any observational strategy for stacked RM should first prioritise extending the spectroscopic baseline, then obtaining more than eight observations for a substantial fraction of the quasars per bin once it has been populated with 500 quasars (there is only a small amount of improvement in lag recovery with the addition of more quasars). The maximum bin width used in the main run as a fraction of the mean luminosity is 0.44 which is for the lowest luminosity bin. In the bin size tests described in Section \ref{sec:binSize} the maximum bin width fraction we tested is 0.4. The fractional redshift bin widths in the binning presented in Figure \ref{fig:lumzdistbins} have a median of 0.13, a minimum of 0.03, and a maximum of 0.53. Luminosity-redshift bins that are at the upper limit of bin width are prone to produce inaccurate lags due to the inclusivity of a broader lag range in the stack. Nevertheless, the number of quasars per bin should not be compromised in favour of a smaller bin width. The entire available photometric light curve should be used to maximise the continuum variability characterisation.

\section{Validation of Stacked RM with Cross-Correlation Functions} \label{sec:CCF}

As previously mentioned, lag recovery methods utilising CCFs have historically been the most popular for RM. The stacked RM studies that have been conducted have all depended on the stacking of CCFs in some way. The debut stacked RM paper by \cite{Fine2012} developed a stacking scheme that directly utilised a basic CCF without any additional technicalities. \cite{Li2018} cleverly innovated a new stacking method that exploited the zDCF. More recently, \cite{Malik2024}, and Elveldt et al. (in prep) stayed close to the method presented by \cite{Fine2013}. The stacking in each of these studies is done by determining an average of the correlation coefficient within a time lag bin which includes all of the light curve data points that fall into that lag bin across all the quasars stacked.

Although we selected {\sc javelin} as the primary lag recovery method for our pipeline, in this section we present an alternative approach using CCFs. This serves as an independent validation of the stacked RM analysis performed with DESI. Repeating the stacked RM analysis using CCFs also provides a consistency check against previous stacked RM studies that employed similar methods. Our cross-correlation analysis follows the methodology outlined in \cite{Fine2012}. The stacking implementation here differs from that of earlier CCF-based RM studies, it more closely resembles the approach we used with the MCMC posterior distributions. Instead of stacking the cross-correlation coefficients directly, we stack the posterior distributions of the cross-correlation function peaks.

\subsection{Adapted Stacked RM method with CCF}

The methodology of producing the mock light curves for the CCF analysis is identical to what has already been explained in detail in Section \ref{sec:mocks_method}. The parameters for these mocks are the same as the ones we used for the main mock run with {\sc javelin}. To summarise, we simulate 250 mock light curves, employing the distributions of flux errors and seasonal gaps that are true to the data, and populating the same luminosity-redshift bins, as displayed in Figures \ref{fig:lumzdistbins}-\ref{fig:errs}. The minimum number of spectroscopic repeats is 2. The photometric baseline for the continuum light curve is set to 1400 days. Instead of feeding the mock light curves into {\sc javelin}, we perform the following CCF analysis.

For each quasar, we simulate 800 light curve realisations. This is inspired by the Monte Carlo technique named `flux randomisation/ random subset selection' (FR/RSS) method developed by \cite{Peterson1998} which is used to generate different realisations of an observed light curve in order to estimate the uncertainty in the CCF lag measurement and test the sensitivity of the lag measurement to the flux errors and sampling characteristics of the light curves. The generating of light curve realisations is used to build up a distribution of the CCF peaks derived from the individual realisations, named the cross-correlation peak distribution (CCPD), and the lag uncertainty is found by measuring its scatter. Since we are dealing with mocks, we have the liberty of simply resampling 800 different observed light curves from the underlying light curve of a particular quasar with different cadences and fluxes to mimic FR/RSS. Usually, a much larger number of realisations are used in the FR/RSS method, On the order of $10^3$ \citep[e.g.,][]{Haas2011} to $10^4$ \citep[e.g.,][]{Chelouche2017}. The justification behind such a large sample size is to statistically remove artefacts in the CCPD related to limitations in the acquiring of data and the method of its processing \citep{Edelson2024}. We find that 800 realisations are enough to distinguish a clear lag peak in the distribution. Furthermore, any artefacts in the individual CCPDs would be suppressed in the stacking procedure. In fact, a stacked CCPD will incorporate a total of 200,000 ($800 \times 250$) realisations which is more than sufficient to minimize the uncertainties associated with small-number statistics. 

For each light curve pair realisation, we calculate a CCF. This is done by calculating the time lag between each continuum and C~{\sc iv} line flux pair, $t_{i,j}= t_ {L,j} -t_{C,i}$. These time lags are used to create 30 equal-population bins of data pairs. This number of bins is chosen to balance resolution in the CCF and maintaining an adequate number of data pairs in each time bin. The correlation coefficient, $r_{k}$, is calculated for the data pairs in each bin, labelled $k$:
\begin{equation} \label{eqn:ccf}
    r_{k} = \frac{1}{n_k}\sum_{i, j}{\frac{(C_i-\bar{C})(L_j-\bar{L})}{\sigma_C\sigma_L}}
\end{equation}
where $C_i$ and $L_j$ are the fluxes of the data points in the continuum and C~{\sc iv} line light curves respectively for the data pairs in bin $k$, $n_k$ is the number of pairs in bin $k$, and $\sigma_c$ and $\sigma_L$ are the root mean squares of the overall continuum and emission line light curves respectively. The CCF is the plot of the correlation coefficient, $r_k$ against the centre lag of time lag bin $k$, $t_k$. Traditional RM would stop here and measure the time lag of a single quasar via the peak or centroid of this function. Previous stacked RM studies like that of \cite{Fine2012} and \cite{Malik2024} would first stack these CCFs with the CCFs of similar quasars before measuring the time lag. We take a different approach here. We calculate the CCF for each realisation of an observed light curve from an objects underlying light curve. The peak of each CCF is used to create a CCPD \citep{Maoz1989}. The CCPD is commonly used to determine the error of the lag measured from the peak of the CCF of the original light curve (e.g., \citealt{Yu2022}). In our case, after examining the CCPDs that are produced by the 800 realisations of each quasar, we found that the peak distributions are very broad and exhibit multiple peaks. This means that basing the lag measurement on the nominal CCF produced from just one light curve realisation of a quasar is unreliable. Therefore, we decided to incorporate the results of the CCPD into the stacked RM lag measurement.
\begin{figure}
\includegraphics{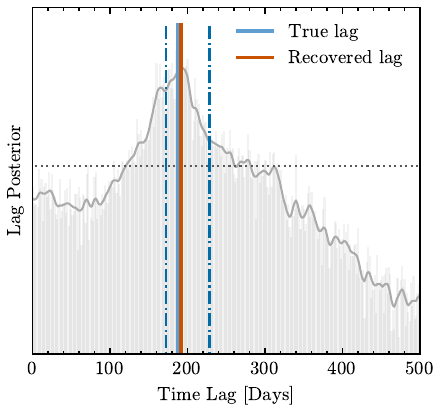} 
\caption{A stacked cross-correlation peak distribution (CCPD) from 250 quasars in a bin with average luminosity at 1350\AA{} of $6.1\times 10^{45} \mathrm{erg\,s^{-1}}$ and average redshift 2.76. The average input lag is 187 days (solid blue line) and the recovered lag is very close at 189 days (solid orange line). The stacked CCPD is purposely shown for a bin which is similar to the bin used to create the stacked lag distribution in Figure \ref{fig:stackedpd} to allow for comparison. The use of CCFs instead of {\sc javelin} resulted in very similar lag distributions qualitatively. The bounds for the peak used in determining the lag uncertainty and SNR are shown as dash-dotted dark-blue lines. The horizontal dark grey dotted line represent a SNR value of one to separate noise from significant peaks.}
\label{fig:stacked_ccpd}
\end{figure}

\begin{figure}
\includegraphics[width=\columnwidth]{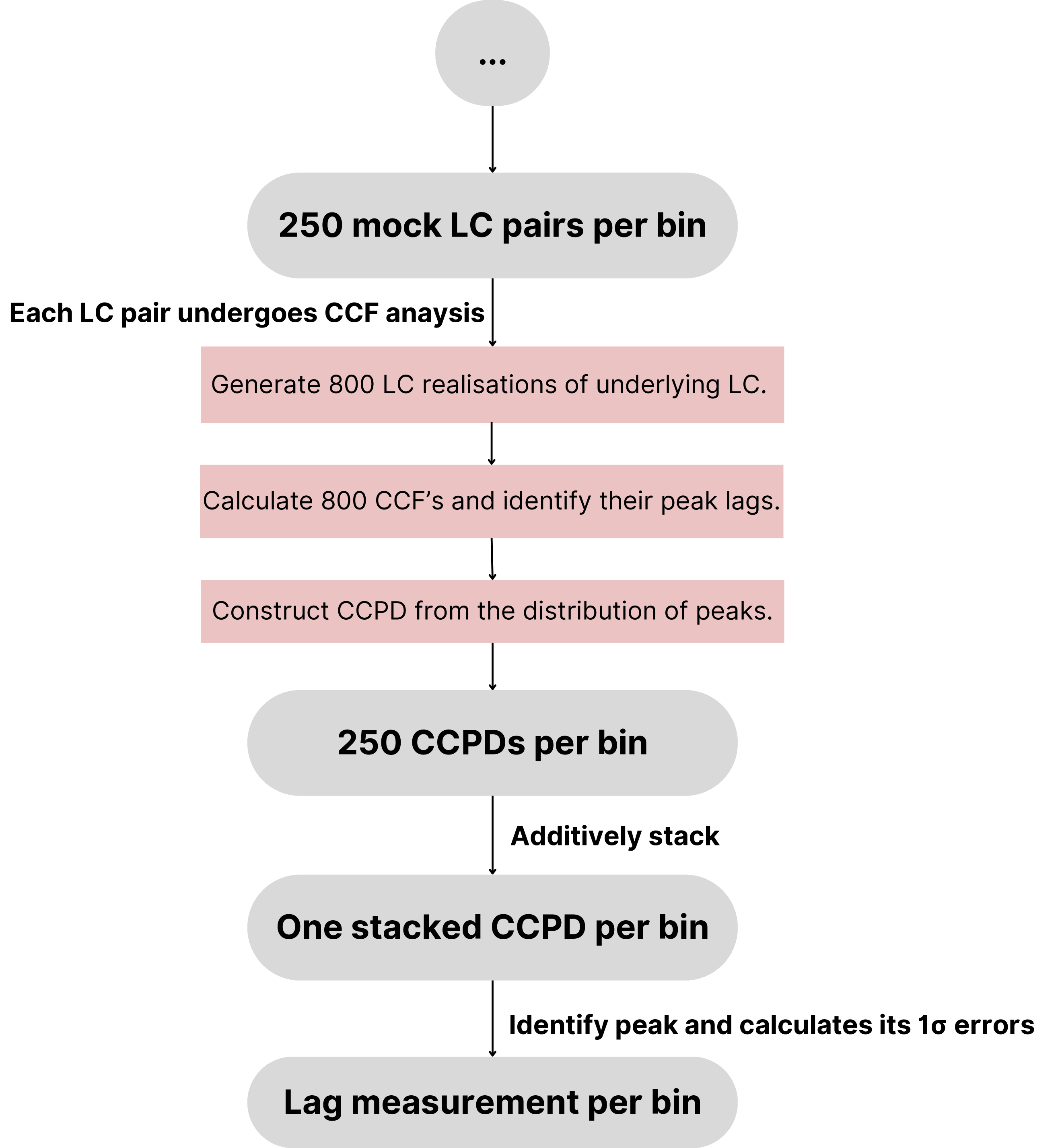} 
\caption{An adapted flow chart to show how stacked RM can be done using cross correlation functions. This first part of the pipeline that produces the mock light curves is identical to Figure \ref{fig:flowchart} and has not been duplicated here.}
\label{fig:flowchart-CCF}
\end{figure}
We stack the CCPDs of each quasar treating it identically to the lag posterior distribution given by {\sc javelin}. In the same way we set a hard lag boundary for the MCMC walkers in {\sc javelin}, we also restrict the peak lags we include in the CCPD to be between 0 and 500 days. The CCPD and {\sc javelin}'s posterior lag distribution are very similar in so far as they are both probability density distributions of the possible lag measurements that can be achieved from one quasar. An example of a stacked CCPD for a bin with average luminosity at 1350\AA{} of $6.1\times 10^{45} \mathrm{erg\,s^{-1}}$ and average redshift 2.76 is shown in Figure~\ref{fig:stacked_ccpd}. This bin of quasars is very similar to the sample of quasars used to build the stacked lag posterior distribution shown in Figure~\ref{fig:stackedpd}, this time with a slightly higher average input lag of 187 days. In both cases, the time lag is well-recovered. Moreover, the SNR of the lag peaks and the overall qualitative structure of the distributions are comparable. Overall, the stacked CCPDs produce lags which are more consistently accurate and have fewer occurrences of outliers. Hence, we confirm in this study, as has been done previously, the ability to recover lags via CCF-based stacked RM. Importantly, we confirm our main pipeline for obtaining lags for DESI quasars with stacked RM since the set-up is identical and all that has been replaced is the lag recovery methodology.

After obtaining the stacked lag CCPD, the method resumes to the main pipeline. To summarise, the stacked CCPD of 250 quasars, each with 800 light curve realisations, is fitted with a Gaussian kernel, the highest lag peak in the distribution is isolated, the location of the MAP is the lag measurement and the 16th/84th percentiles of the peak are the errors for this luminosity-redshift bin.
We present in Figure \ref{fig:flowchart-CCF} the alternative latter half of the stacked RM pipeline with CCFs which serves as an independent check to the {\sc javelin} pipeline show in Figure \ref{fig:flowchart}.

\subsection{Recovery of the $R$-$L$ relationship with CCF}
We provide here a corresponding set of results to those of the main pipeline for the sake of comparison.
Figure \ref{fig:rlz_ccf} presents the recovered lags for all the bins on the $R$-$L$ plane. Overall, the CCF recovered lags have a smaller uncertainty than the {\sc javelin} pipeline as can be seen by the generally smaller error bars. At the higher luminosity end the results are very similar to the {\sc javelin} pipeline. However, there is a clear contrast at the lower luminosity end. The pipeline with CCFs seems to struggle to recover low luminosity lags as seen by their deviation from the input \cite{Hoormann2019} relation and the inflation of the error bars. There is a bias towards underestimation of the lag. This seems to be an exacerbation of what has already been found in the {\sc javelin} pipeline. The underestimation is a feature of the MCMC lag posteriors and this bias is even more exaggerated in our analysis with CCFs. In both the {\sc javelin} and CCF recovered $R$-$L$s, the underestimation is seen to be more pronounced at low luminosities. This confirms that the bias is method independent. Our results reveal that the CCF-substituted pipeline shows a higher level of sensitivity to the undetermined cause of systematic underestimation.
\begin{figure}
\includegraphics{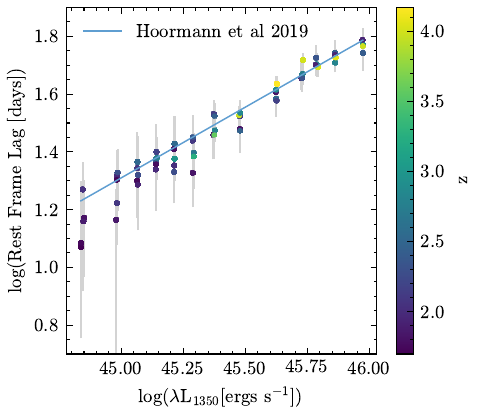} 
\caption{The recovered lag of each stacked CCPD plotted on the radius-luminosity plane. Details same as Figure \ref{fig:rlz}. The results suggest that, for these idealised survey simulations, CCFs can successfully recover time lags through stacked RM despite sparse spectroscopic cadence. However, this performance does not generalise to cases with aligned seasonal gaps between spectroscopy and photometry, where the CCF lag recovery deteriorates substantially.}
\label{fig:rlz_ccf}
\end{figure}

Figure \ref{fig:snr_ccf} presenting the SNR of the lag peaks in the CCPDs depicts the same trends we have seen before in relation to luminosity and redshift. There is an inverse relation between both luminosity and redshift and the SNR. Fascinatingly, the overall SNR ratio with the CCF method is higher than for the main pipeline. For reference, the lag peak in the stacked posterior distributions has SNRs ranging from 1.05 to 1.45 whilst the peak in the stacked CCPDs has SNRs ranging between 1.24 and 2.11. Visual inspection of the stacked CCPDs makes it evident what the reason for this is: the stacked CCPDs have a very low rate of secondary false peaks compared to {\sc javelin}'s stacked lag distributions. In other words, the stacked CCPDs appeared to have a higher level of smoothness so the `noise' is weaker and has less power to diminish the signal. This can even be seen in the examples already presented in Figure \ref{fig:stackedpd} and Figure \ref{fig:stacked_ccpd}. Whereas {\sc javelin} is plagued with numerical issues involving walkers getting stuck at local minima and forging aliases, CCF calculations evades these complications.

\begin{figure}
\includegraphics{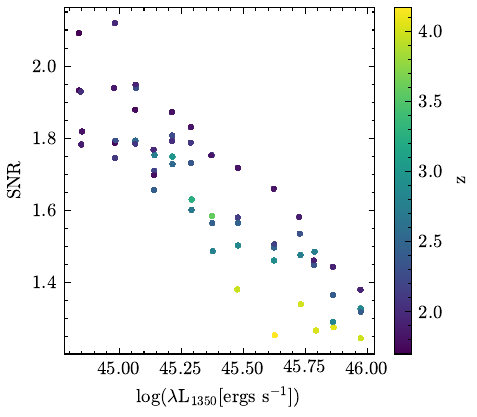} 
\caption{The SNR of the CCPD lag peaks for each luminosity-redshift bin. Details same as figure \ref{fig:snr-L}.}
\label{fig:snr_ccf}
\end{figure}

The recovered $R$-$L$ relation from the CCF pipeline is shown in Figure \ref{fig:rl_ccf} which has the form:
\begin{equation}
\log R \text{[days]} =  (0.73\pm 0.023) + (0.54 \pm 0.017)\log\frac{\lambda L_\lambda(1350\text{\AA{}})}{10^{44}[\mathrm{erg\,s^{-1}}]}.
\label{eqn:recoveredRL}
\end{equation}
This $R$-$L$ relation does not agree with the input $R$-$L$ given in Equation \ref{eqn:HoormannRL} to 1$\sigma$. It is actually only in agreement within 4$\sigma$ and this is due to the pronounced underestimation of lags at low luminosities combined with smaller uncertainties. 
Although CCPD lags appear to be more precise than those from {\sc javelin}, they exhibit a stronger systematic bias toward underestimation. As a result, the final $R$-$L$ relation derived from CCPD lags is more susceptible to being offset toward lower lag or radius values, with uncertainties that may give a misleading impression of accuracy. This finding is present in the literature which led to our selection of {\sc javelin} for the main pipeline (see Section \ref{sec:javelin}). \citet{KilerciEser2015} reported that their $R$-$L$ fit for the CCF lags had a steeper slope than that from {\sc javelin} lags. This is also what has been found in this study, a slope of 0.51 for {\sc javelin} compared to a slope of 0.54 for CCF. 

That said, the systematic uncertainties and biases for {\sc javelin} and CCFs are both small and within the intrinsic scatter of the observed $R$-$L$ relationship. The {\sc javelin} lags had a scatter of 0.033 dex and the CCF lags had a scatter of 0.024 dex both measured around the input $R$-$L$ relation. These are both less than $5\%$ of the observed scatter in the C~{\sc iv} $R$-$L$ relationship ($\sim$0.5 dex, \citealt{Shen2024}). Despite the fact that CCF lags seemed to be more precise in the main run they were still less accurate compared to {\sc javelin}, and so this must be taken into account if they are used in future stacked RM studies. It may be most suitable to restrict the use of CCFs for stacked RM with DESI-like data to only the high-luminosity regime, in order to circumvent the effects present at low luminosities. Beyond the findings of the main run, additional tests determining lags from CCPDs using mock light curves with degraded cadences and amplified seasonal gaps, thereby further reducing data quality, proved to be detrimental to the performance of the CCF-based approach. Seasonal aliasing came to dominate the CCPD shape and consequently produced null results. Since CCPD lag recovery completely malfunctions when seasonal gaps are enhanced, their application is thus restricted.

\begin{figure}
\includegraphics{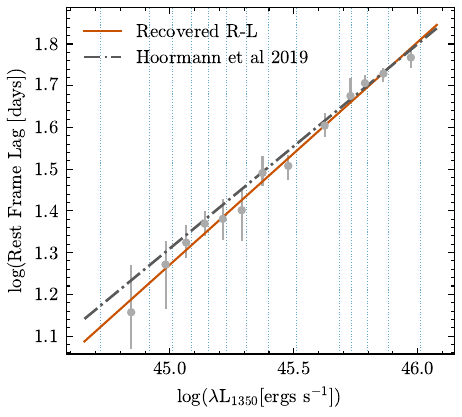} 
\caption{The recovered radius-luminosity relation using CCFs shown as a solid red line. Details same as Figure \ref{fig:rl}.}
\label{fig:rl_ccf}
\end{figure}

\section{Conclusion} \label{sec:conc}
The results of this feasibility study conclude that stacked RM with DESI-like data is possible using the Bayesian MCMC code {\sc javelin}. This has been proved with a mock quasar light curve sample that was simulated using a DRW model and sampled according to DESI and ZTF observational properties. The lag between the continuum and C~{\sc iv} broad emission line light curves was set according to the C~{\sc iv} $R$-$L$ relation found in \citet{Hoormann2019}. The stacked RM method described in this paper recovered the input $R$-$L$ relation within 1$\sigma$. This demonstrates that a survey designed for cosmology may be used for RM to reach a better understanding of quasar populations. The power to measure lags degrades with higher luminosity quasars due to lower variability light curves. Future work is needed to strengthen the robustness of this regime. We identified several factors that may effect the measurement of a time lag and we tested the robustness of the pipeline against these, the conclusions of which are:
\begin{itemize}
    \item Inflating the uncertainty in the broad emission line flux measurement can degrade the accuracy of the recovered lag, however, the overall impact is small since the $R$-$L$ relation remains to be recovered to 1$\sigma$ (see Section \ref{sec:flux_err}).
    \item Choosing photometric surveys that have a multi-year baseline (around $\gtrsim$1000 days) is also strongly recommended for sufficient continuum variation characterisation (see Section \ref{sec:phot_compl}).
    \item We recommend stacking at least 400 objects in each luminosity-redshift bin to minimise the lag uncertainty (see Section \ref{sec:numStack}).
    \item Quasars with $\geq 2$ spectroscopic observations can be included in the stack. Limiting the sample to quasars with more spectroscopic epochs (>2) is not justifiable considering the weak improvement in lag recovery (see Section \ref{sec:extra}). Rather, it is highly beneficial to focus on lengthening the baseline of the spectroscopic light curves instead.
\end{itemize}
The stacked RM results were validated by conducting an alternative stacked RM method with CCFs which confirmed that the time lags could be recovered. However, CCFs are affected by systematic uncertainties to a greater degree than {\sc javelin} whilst at the same time giving smaller errors, therefore rendering the lags unreliable. The CCF-based stacked RM method also fails to recover lags when the light curves are degraded, particularly when seasonal gaps are amplified.

\section*{Data Availability}
The DESI data at the foundation of this study is publicly available as part of Data Release 1 (see \url{https://data.desi.lbl.gov/doc/releases/dr1/}). The data that can be used to recreate all the figures in this paper is made public in a Zenodo repository.

\section*{Acknowledgements}
Many thanks go to the referee for comments that improved the quality of this manuscript. Special thanks to the Galaxy \& Quasar Physics working group in DESI for discussion and contribution to many aspects of this project. We thank Andrew Penton and Natalie Eir\'{e} Sommer for authoring and making available the mock LC simulation code used in this study. 

RA is supported by a UK Science and Technology Facilities Council (STFC) studentship and the University of Portsmouth. EM acknowledges support from an STFC Ernest Rutherford Fellowship, with grant reference ST/W004755/1. 

This material is based upon work supported by the U.S. Department of Energy (DOE), Office of Science, Office of High-Energy Physics, under Contract No. DE–AC02–05CH11231, and by the National Energy Research Scientific Computing Center, a DOE Office of Science User Facility under the same contract. Additional support for DESI was provided by the U.S. National Science Foundation (NSF), Division of Astronomical Sciences under Contract No. AST-0950945 to the NSF’s National Optical-Infrared Astronomy Research Laboratory; the Science and Technology Facilities Council of the United Kingdom; the Gordon and Betty Moore Foundation; the Heising-Simons Foundation; the French Alternative Energies and Atomic Energy Commission (CEA); the National Council of Humanities, Science and Technology of Mexico (CONAHCYT); the Ministry of Science, Innovation and Universities of Spain (MICIU/AEI/10.13039/501100011033), and by the DESI Member Institutions: \url{https://www.desi.lbl.gov/collaborating-institutions}. Any opinions, findings, and conclusions or recommendations expressed in this material are those of the author(s) and do not necessarily reflect the views of the U. S. National Science Foundation, the U. S. Department of Energy, or any of the listed funding agencies.

The authors are honoured to be permitted to conduct scientific research on I'oligam Du'ag (Kitt Peak), a mountain with particular significance to the Tohono O’odham Nation.


\bibliographystyle{mnras}



\appendix

\section{Dependency on the time lag prior} \label{app:prior}

\begin{figure}
\includegraphics{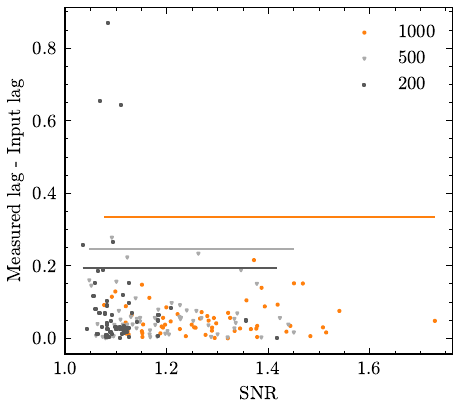}
\caption{The difference between the measured log lag and the input log lag of a bin against the SNR of the lag peak in the bins stacked posterior distribution. The results are shown for three different runs where the only variable changed is the width of the lag prior in {\sc javelin}: a fixed range 0-1000 days (orange), a fixed range 0-500 days (light gray), and a relative range of 200 days centred around the expected lag in that bin (dark grey). The horizontal line is situated at the average deviation of the measured lag from the input lag and spans the range of SNRs in the run. The accuracy of the lags recovered is increased with tighter lag priors. Counterintuitively, the SNR range shifts to lower values with tighter priors. This is because when we reduce the lag range to be around the expected lag the MCMC walkers are more likely to explore the entire range more equally compared to a very wide range where higher lags are not probed as much by the walkers making the lag peak stand out from the rest of the distribution.}
\label{fig:priors}
\end{figure}

{\sc javelin} allows one to define lag limits when calculating the posterior distributions. The chosen prior may influence the final lag measured. We tested three prior lag ranges: 
\begin{itemize}
    \item A fixed range of [0,500] to cover all the expected lags for the entire simulation suite.
    \item A fixed range of [0,1000] to reflect how finding longer lags is possible given the photometric light curve length available, as well as the 8-year DESI survey length. This lag range prior has the least bias towards the expected results.
    \item A relative physical range [$l$-100, $l$+100] where $l$ is the expected average lag for that luminosity-redshift bin. This range has a width of 200 days. This may bias the recovered lag. Avoiding the use of a previously determined $R$-$L$ relation to limit prior assumptions is preferred, since we want our method to produce an independent $R$-$L$ relation.
\end{itemize}
Usually, a negative lag range is included to estimate the occurrence rate of false positives since negative lags are unphysical according to our current understanding of AGN geometry and kinematics. In this study, we have not opened the prior space to negative lags as we are working with simulations with known lags and therefore can directly deduce which lags are false detections. Also, the continuum light curve is truncated 50 days after the date of the last spectroscopic epoch so a negative lag range cannot be properly explored due to the light curve set up.

We found that the smaller the prior lag range, the more accurate the lags recovered, and hence a better recoverability of the $R$-$L$ relation. These results, shown in Figure \ref{fig:priors}, indicate it is important to carefully pick the prior lag range due to the sensitivity of {\sc javelin}'s MCMC method. Providing a wider prior counter-intuitively results in a stronger lag signal (higher SNR). Due to the sparse light curves, the MCMC walkers explore the whole range of lags. A larger range dilutes this "exploration" noise in the posterior distribution and allows the correct lag peak to stand out more. In a tight prior the walkers explore a more concentrated space so the lag peak and noise do not have as much contrast in the posterior distribution.

\section{Effects of the presence of seasonal gaps} 
\label{app:lagsgaps}

Part of our initial testing of the pipeline was to see if a null result is recovered with the removal of the lag between the continuum and emission line light curves and, further, to what extent does the absence of seasonal gaps enhance the lag result. 

We conducted tests by modifying the suite of light curve pairs so that there is no lag between them and by completely removing seasonal gaps in their sampling. We conducted four pipeline runs across all bins for the following test cases: no lags and no seasonal gaps, with lags and without seasonal gaps, no lags with seasonal gaps, and with both lags and seasonal gaps.

Figure \ref{fig:lg_stacked} shows the average stacked lag distributions across all bins for each run. When the lag is removed we find that on average we achieve a null result, with the peak of the averaged posterior distributions being at zero. Importantly, these distributions show that no systematic aliases, or significant secondary peaks, appear when there are seasonal gaps present in the mock photometric light curves. There is an upturn at the upper lag boundary which appears due to {\sc javelin}'s oversampling near the prior boundary as reported by \citet{Mcdougall2025}. This is more noticeable in light curves without seasonal gaps. What is clear from these tests is that seasonal gaps considerably diminish the lag peak which can be seen in both the runs with and without lags. This illustrates the observational limitations of RM since seasonal gaps restrict the SNR achievable by stacking. The position of lag peaks in both lag runs are in agreement. The average simulated lag across all bins is 111 days in both runs. The results of the two runs both recover an average lag of 109 days which is within 1$\sigma$ of the average simulated lag. We would expect the MAP of the averaged distribution across all bins to fall on 109 days. Surprisingly, it is found at a much lower lag around 60 days. This provides a hint towards the systematic underestimation of lags produced by the pipeline. Although lags are mostly recovered well in individual bins, meaning the MAP of the peak is at the correct lag, the averaging of the stacked posterior distribution reveals the bias towards lower lags across all stacked distributions.

\begin{figure}
\includegraphics{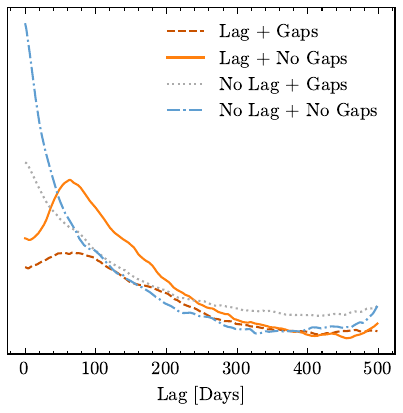}

\caption{Average of stacked lag posterior distributions of all bins for four runs with modified light curves. The peak of the runs with no lag simulated between the light curve pairs are at zero as expected with no additional aliases due to observational cadence. The peak of the distributions with a lag roughly corresponds to the median lag over all luminosities and redshifts. This shows that no additional aliasing appears with the presence of seasonal gaps. However, the significance of the peak does drop significantly.}
\label{fig:lg_stacked}
\end{figure}

\begin{figure}
\includegraphics{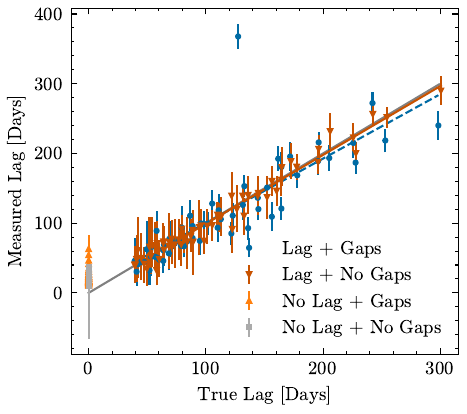}
\caption{Measured versus simulated lags for each bin in four runs with modified light curves. The scatter around the ratio of equality (solid dark grey line) increases with the presence of seasonal gaps for both runs with a lag and no lag. The line of best fit for the points with both lag and gaps (dashed blue line) shows that there is a slight underestimation in lag measurement caused by the presence of seasonal gaps at the high lag end, corresponding to higher luminosities. The solid red line is almost coincident with the line of equality presenting the power of this method in the absence of seasonal gaps.}
\label{fig:lg_lags}
\end{figure}

Figure \ref{fig:lg_lags} shows the measured lag compared to the true lag for all bins in each run. The removal of seasonal gaps reduced the scatter in the lag recovery. For the run without a lag, we see more points clustered towards smaller lags, closer to the true lag zero, when there are no seasonal gaps. In both cases, the scatter remains below 50 days.  For the runs with a lag, the average scatter is 18 days when seasonal gaps are included, and it is reduced by half, to 9 days, once these gaps are removed. These runs give us an estimation of the expected false positive rate. A clear example is the extreme outlier in the run with both lags and gaps (blue point towards the top of the plot), which occurred since there is a secondary false peak with a similar probability as the true lag peak in the distribution. These outliers happen spuriously and have a low occurrence rate. In the runs with a lag, seasonal gaps reduce lag recovery, especially at longer lags. The magnitude in the scatter of the lag runs directly reflect the magnitude of scatter in the no lag runs, with and without gaps respectively.

\vspace{10mm}

\textit{ \small
$^{1}$Institute of Cosmology \& Gravitation, University of Portsmouth, Dennis Sciama Building, Burnaby Road, Portsmouth, PO1 3FX, UK\\
$^{2}$Department of Physics and Astronomy, University of Sussex, Brighton BN1 9QH, UK\\
$^{3}$Lawrence Berkeley National Laboratory, 1 Cyclotron Road, Berkeley, CA 94720, USA \\
$^{4}$Department of Physics, Boston University, 590 Commonwealth Avenue, Boston, MA 02215 USA \\
$^{5}$Centre for Extragalactic Astronomy, Department of Physics, Durham University, South Road, Durham, DH1 3LE, UK\\
$^{6}$Dipartimento di Fisica “Aldo Pontremoli”, Universit`a degli Studi di Milano, Via Celoria 16, I-20133 Milano, Italy \\
$^{7}$Department of Physics \& Astronomy, University College London, Gower Street, London, WC1E 6BT, UK \\
$^{8}$School of Physics and Astronomy, University of Southampton, Southampton, SO17 1BJ, UK \\
$^{9}$School of Mathematics and Physics, University of Queensland, Brisbane, QLD 4072, Australia\\
$^{10}$Instituto de F\'{\i}sica, Universidad Nacional Aut\'{o}noma de M\'{e}xico,  Circuito de la Investigaci\'{o}n Cient\'{\i}fica, Ciudad Universitaria, Cd. de M\'{e}xico  C.~P.~04510,  M\'{e}xico \\
$^{11}$School of Mathematics, Statistics and Physics, Newcastle University, Newcastle upon Tyne, NE1 7RU, UK\\
$^{12}$Institut de F\'{i}sica d'Altes Energies (IFAE), The Barcelona Institute of Science and Technology, Edifici Cn, Campus UAB, 08193, Bellaterra (Barcelona), Spain \\
$^{13}$Departamento de F\'isica, Universidad de los Andes, Cra. 1 No. 18A-10, Edificio Ip, CP 111711, Bogot\'a, Colombia \\
$^{14}$National Astronomical Observatories, Chinese Academy of Sciences, A20 Datun Road, Chaoyang District, Beijing, 100101, China\\
$^{15}$Fermi National Accelerator Laboratory, PO Box 500, Batavia, IL 60510, USA \\
$^{16}$Center for Cosmology and AstroParticle Physics, The Ohio State University, 191 West Woodruff Avenue, Columbus, OH 43210, USA\\
$^{17}$NSF NOIRLab, 950 N. Cherry Ave., Tucson, AZ 85719, USA\\
$^{18}$Department of Physics and Astronomy, University of California, Irvine, 92697, USA \\
$^{19}$Center for Astrophysics|Harvard \& Smithsonian, 60 Garden Street, Cambridge, MA 02138, USA \\
$^{20}$Sorbonne Universit\'{e}, CNRS/IN2P3, Laboratoire de Physique Nucl\'{e}aire et de Hautes Energies (LPNHE), FR-75005 Paris, France \\
$^{21}$Department of Physics and Astronomy, Siena University, 515 Loudon Road, Loudonville, NY 12211, USA\\
$^{22}$Department of Astronomy, University of Illinois at Urbana-Champaign, Urbana, IL 61801, USA\\
$^{23}$International Gemini Observatory/NSF NOIRLab, Casilla 603, La Serena, Chile \\
$^{24}$Departament de F\'isica, EEBE, Universitat Polit\`ecnica de Catalunya, c/Eduard Maristany 10, 08930 Barcelona, Spain\\
$^{25}$Instituto de Astrof\'{i}sica de Andaluc\'{i}a (CSIC), Glorieta de la Astronom\'{i}a, s/n, E-18008 Granada, Spain \\
$^{26}$Department of Physics and Astronomy, The University of Utah, 115 South 1400 East, Salt Lake City, UT 84112, USA \\
$^{27}$Department of Physics and Astronomy, Sejong University, 209 Neungdong-ro, Gwangjin-gu, Seoul 05006, Republic of Korea \\
$^{28}$CIEMAT, Avenida Complutense 40, E-28040 Madrid, Spain\\
$^{29}$Department of Physics, University of Michigan, 450 Church Street, Ann Arbor, MI 48109, USA\\
$^{30}$Institute of Space Sciences, ICE-CSIC, Campus UAB, Carrer de Can Magrans s/n, 08913 Bellaterra, Barcelona, Spain\\
$^{31}$National Astronomical Observatories, Chinese Academy of Sciences, A20 Datun Road, Chaoyang District, Beijing, 100101, P.~R.~China
}


\bsp	
\label{lastpage}
\end{document}